\documentclass[showkeys,superscriptaddress,floatfix,aps,10pt,prd,twocolumn]{revtex4-2}
\usepackage{graphicx,epstopdf}
\usepackage[caption=false]{subfig}
\usepackage{appendix}
\usepackage[T1]{fontenc}
\usepackage{lmodern}
\usepackage[dvipsnames,x11names]{xcolor}
\usepackage[colorlinks=true,linkcolor=ForestGreen,citecolor=ForestGreen,urlcolor=NavyBlue]{hyperref}
\usepackage[sort&compress]{natbib}
\usepackage{morefloats}
\usepackage[pdf]{pstricks}
\usepackage{amsmath}
\usepackage{amssymb}
\usepackage{amsfonts}
\usepackage{rotating}
\usepackage{cancel}
\usepackage{mathtools}
\usepackage{bbm}
\usepackage{dsfont}
\usepackage{bbold}
\usepackage{multirow}
\usepackage{ulem}
\usepackage{physics}
\usepackage{orcidlink}
\usepackage{colortbl}

\def\xslash{x\!\!\!\slash }

\def\vel{\left|}
\def\ver{\right|}

\begin{document}

\title{Investigating the underlying structure of vector hidden-charm tetraquark states via their electromagnetic characteristics}

\author{Ula\c{s}~\"{O}zdem\orcidlink{0000-0002-1907-2894}}%
\email[]{ulasozdem@aydin.edu.tr }
\affiliation{ Health Services Vocational School of Higher Education, Istanbul Aydin University, Sefakoy-Kucukcekmece, 34295 Istanbul, T\"{u}rkiye}

 
\begin{abstract}
Accessing a full picture of the internal structure of hadrons would be a key topic of hadron physics, with the main motivation to study the strong interaction binding the visible matter.  Furthermore, the underlying structure of known exotic states remains an unresolved fundamental issue in hadron physics, which is currently being addressed by hadron physics community. It is well known that electromagnetic characteristics can serve as a distinguishing feature for states whose internal structures are complex and not yet fully understood. The aim of this study is to determine the magnetic moments of vector hidden-charm tetraquark states by making use of QCD light-cone sum rules. In order to achieve this objective, the states mentioned above are considered in terms of the diquark-antidiquark structure. Subsequently, a comprehensive examination is conducted, with four distinct interpolating currents being given particular consideration, as these have the potential to couple with the aforementioned states.   It has been observed that there are considerable discrepancies between the magnetic moment results extracted employing different diquark-antidiquark structures. Such a prediction may be interpreted as the possibility of more than one tetraquark with the identical quantum numbers and similar quark constituents, but with different magnetic moments. The numerical predictions yielded have led to the conclusion that the magnetic moments of the vector hidden-charm tetraquark states are capable of projecting the inner structure of these states, which may then be used to determine their quark-gluon structure and quantum numbers. In order to provide a comprehensive analysis, the individual quark contributions to the magnetic moments are also examined. As a byproduct, the quadrupole moments of these states are also extracted, and the obtained values differ from zero, which indicates that the charge distribution of these states is not spherical. 
\end{abstract}

\maketitle

\section{motivation}\label{motivation}

Over the past two decades, hadron physics has witnessed a period of significant advancement. This growth can be attributed to the accumulation of valuable data on the properties of hadrons, as well as the emergence of novel theoretical insights and predictions that have expanded our understanding of the quark-gluon structure of the fundamental building blocks of matter. One of the most intriguing findings among these accomplishments is the observation of hadrons that can be interpreted as four- and five-quark states. 
In 2003, the Belle Collaboration made an experimental discovery of states that may contain four quarks, namely the observation of the X(3872) state~\cite{Belle:2003nnu}. Following this initial discovery, a considerable number of new states of possible four- or five-quark structures have been observed by different experimental collaborations. As new observations are made, the field of exotic states continues to expand, representing a dynamic and evolving area of hadron physics. Different theoretical models have been proposed to elucidate the true nature of these states. In the case of tetraquarks, the most widely accepted interpretations of their internal structures are the molecular and compact diquark-antidiquark configurations. However, it remains challenging to accurately specify the inner structures of these states. 
A list of the most recent advances in the domain of exotic states can be found in Refs.~\cite{Esposito:2014rxa, Esposito:2016noz,Olsen:2017bmm, Lebed:2016hpi,Nielsen:2009uh, Brambilla:2019esw,Agaev:2020zad, Chen:2016qju,Ali:2017jda, Guo:2017jvc,Liu:2019zoy, Yang:2020atz,Dong:2021juy, Dong:2021bvy, Meng:2022ozq, Chen:2022asf}.

In recent years, some vector hidden-charmed tetraquark states, including Y(4220/4260), Y(4360/4390), and Y(4630/4660), have been observed. However, these states cannot be well correlated with the standard meson with quantum numbers $J^{PC}$ = $1^{--}$. In order to gain insight into the nature of these states, a multitude of theoretical models have been put forth, and extensive research has been conducted on them. In addition to the aforementioned discovered states, there may also exist vector hidden-charm tetraquark states with disparate quark contents and structures. 
In Ref.~\cite{Ozdem:2022kck}, we conducted a comprehensive analysis of the magnetic moments of several experimentally observed vector hidden-charm states within the framework of QCD light-cone sum rules and considered the compact diquark-antidiquark structure.  In this study, we extend our analysis to calculate the magnetic moments of other potential vector hidden-charm tetraquarks with and without strangeness using the QCD light-cone sum rule method~\cite{Chernyak:1990ag, Braun:1988qv, Balitsky:1989ry} and assume that they have a diquark-antidiquark internal configuration.
In studying the electromagnetic characteristics of hadrons, apart from their spectroscopic properties, researchers can gain insight into the nature of these particles, their internal structure, and their quantum numbers. It is established that the electromagnetic characteristics of hadrons, particularly their magnetic moments, which pertain to the spatial distribution of charge and magnetization within the hadrons, are associated with the spatial distribution of quarks and gluons within them. The analysis of the magnetic and higher multipole moments of hadrons represents a compelling avenue of research.  The existing body of literature comprises a plethora of studies that examine the electromagnetic characteristics of both hidden- and doubly-heavy tetraquark states. These studies endeavor to elucidate both the internal structure of these states and their true nature~\cite{Ozdem:2024dbq, Ozdem:2024lpk, Mutuk:2023oyz, Wang:2023bek, Ozdem:2023rkx, Ozdem:2023frj, Lei:2023ttd, Zhang:2021yul, Azizi:2023gzv, Ozdem:2022eds, Ozdem:2022kck, Xu:2020qtg, Wang:2017dce, Ozdem:2022yhi, Wang:2023vtx, Ozdem:2021hmk, Azizi:2021aib, Ozdem:2021hka, Xu:2020evn, Ozdem:2021yvo, Ozdem:2017exj, Ozdem:2017jqh, Ozdem:2024rrg, Mutuk:2024vzv, Ozdem:2024jmb}. Nevertheless, it is apparent that additional research is necessary to obtain a more profound comprehension of the electromagnetic properties of states with diverse structural compositions.

The remainder of this study is structured as follows: Section~\ref{formalism} presents the mathematical formalism of QCD light-cone sum rules, which are employed to determine an adequate correlation function within both the hadronic and QCD domains. The required sum rules for the magnetic moments are derived following the established methodology. Section~\ref{numerical} is devoted to the numerical analysis of the magnetic moments. Section~\ref{sum} comprises the study's summary and outlook.

 \begin{widetext}
 
\section{The method}\label{formalism}

This section of the manuscript presents the derivation of QCD light-cone sum rules for the purpose of obtaining the magnetic moments of vector hidden-charm ($Y_{c \bar c}$ for short) tetraquark states. To derive the relevant formalism, it is necessary to consider the following correlation function, which represents the key parameter of the method. 
\begin{equation}
 \label{edmn01}
\Pi _{\alpha \beta }(p,q)=i\int d^{4}x\,e^{ip\cdot x}\langle 0|\mathcal{T}\{J^i_{\alpha}(x)
J_{\beta }^{i \dagger }(0)\}|0\rangle_{F}, 
\end{equation}
where $F$ denotes the external electromagnetic field, while $J^i_{\alpha(\beta)}(x)$ denotes the interpolating currents, which correspond to the vector hidden-charm tetraquark states. 
In this study, we posit that the vector hidden-charm tetraquark states are constituted by $C \gamma_5-\gamma_5 \gamma_\alpha C$ and $C-\gamma_\alpha C$-type diquark structures. Accordingly, we construct the interpolating currents as follows:
\begin{align}
\label{curr1}
J_{\alpha }^{1}(x) &= \varepsilon^{abc} \varepsilon^{ade}  \big[q_1^{b^T} (x) C \gamma_5 c^c (x)\big]\big[\bar c^d (x) \gamma_5 \gamma_\alpha C \bar q_2^{e^T} (x) \big],\\
J_{\alpha }^{2}(x) &= \varepsilon^{abc} \varepsilon^{ade} \big[q_1^{b^T} (x) C \gamma_\alpha \gamma_5   c^c (x)\big]\big[\bar c^d (x) \gamma_5 C \bar q_2^{e^T} (x) \big],\\
J_{\alpha }^{3}(x) &= \varepsilon^{abc} \varepsilon^{ade}  \big[q_1^{b^T} (x) C  c^c (x)\big]\big[\bar c^d (x)  \gamma_\alpha C \bar q_2^{e^T} (x) \big],\\
J_{\alpha }^{4}(x) &= \varepsilon^{abc} \varepsilon^{ade} \big[q_1^{b^T} (x) C \gamma_\alpha c^c (x)\big]\big[\bar c^d (x) C \bar q_2^{e^T} (x) \big],
\label{curr4}
\end{align}
where $a$, $b$, $c$, $d$ and $e$ are color indices, and $C$ is the charge conjugation operator. The quark content of the vector hidden-charm tetraquark states is presented in Table \ref{quarkcon}. It is important to note that, given that these interpolating currents possess identical quantum numbers, it is reasonable to conclude that they would couple to the same tetraquark states.
\begin{table}[htp]
	\addtolength{\tabcolsep}{10pt}
		\begin{center}
		\caption{The quark content of the vector hidden-charm tetraquark states.}
	\label{quarkcon}
\begin{tabular}{lccccccccc}
	   \hline\hline
	   \\
  Quarks& $[uc][\bar c \bar d]$&$[uc][\bar c \bar s]$& $[dc][\bar c \bar s]$& $[sc][\bar c \bar s]$\\
  \\
\hline\hline
$q_1$&  u & u & d & s \\
$q_2$&  d & s & s &s \\
	   \hline\hline
\end{tabular}
\end{center}
\end{table}

Following the principles of the QCD light-cone sum rules approach, the analytical procedures can be formulated in the following way:
\begin{itemize}
 \item The correlation function is formulated regarding hadronic quantities such as mass, magnetic moment, form factor, etc., is so-called "hadronic description".
 
 \item The correlation function is also formulated regarding quantities associated with quark-gluon degrees of freedom and distribution amplitudes, etc., and is called the "QCD description".
 
 \item In the last step, the above-mentioned notations are matched using the assumption of quark-hadron duality. In order to exclude any potential inclusions that may be deemed undesirable from the calculations, a double Borel transformation and continuum subtractions are then carried out in order to derive the sum rules for the desired physical quantity, which is to be determined.
\end{itemize}

\subsubsection{Hadronic Description}

To derive the hadronic description of the correlation function, we insert a complete set of intermediate vector hidden-charm tetraquark states with the same quantum numbers as the interpolating currents into the correlation function and then perform the integral over x. The resulting expression is as follows:
\begin{align}
\label{edmn04}
\Pi_{\alpha\beta}^{Had} (p,q) &= {\frac{\langle 0 \mid J_\alpha (x) \mid
Y_{ c \bar c}(p,\varepsilon^i) \rangle}{p^2 - m_{Y_{ c \bar c}}^2}} \langle Y_{ c \bar c} (p, \varepsilon^i) \mid Y_{ c \bar c} (p+q, \varepsilon^f) \rangle_F 
\frac{\langle Y_{ c \bar c} (p+q, \varepsilon^f) \mid J_{\beta }^{\dagger } (0) \mid 0 \rangle}{(p+q)^2 - m_{Y_{ c \bar c}}^2}+ \cdots,
\end{align}
where $\cdots$ stands for continuum and higher states.

As can be seen from Eq. (\ref{edmn04}), the explicit forms of certain matrix elements are needed. These are presented below:
\begin{align}
\label{edmn05}
\langle 0 \mid J_\alpha (x) \mid Y_{ c \bar c} (p, \varepsilon^i) \rangle &=  \lambda_{Y_{ c \bar c}} \varepsilon_\alpha^i\,,\\
\langle Y_{ c \bar c} (p+q, \varepsilon^{f}) \mid J_{\beta }^{\dagger } (0) \mid 0 \rangle &= \lambda_{Y_{ c \bar c}} \varepsilon_\beta^{* f}\,,\\
\langle Y_{ c \bar c}(p,\varepsilon^i) \mid  Y_{ c \bar c} (p+q,\varepsilon^{f})\rangle_F &= - \varepsilon^\gamma (\varepsilon^{i})^\mu (\varepsilon^{f})^\nu
\Big[ G_1(Q^2)~ (2p+q)_\gamma ~g_{\mu\nu}  
+ G_2(Q^2)~ ( g_{\gamma\nu}~ q_\mu -  g_{\gamma\mu}~ q_\nu)
\nonumber\\ 
&
- \frac{1}{2 m_{Y_{ c \bar c}}^2} G_3(Q^2)~ (2p+q)_\gamma
q_\mu q_\nu  \Big],\label{edmn06}
\end{align}
where $\lambda_{Y_{ c \bar c}}$ is the pole residue,  $ \varepsilon_\alpha^i $ and $\varepsilon_\beta^{*f}$ are the polarization vector of the initial and final $Y_{ c \bar c}$ tetraquark states, respectively; $\varepsilon^\gamma$ is the polarization vector of the photon, and  $G_i(Q^2)$'s are the corresponding radiative transition form factors with  $Q^2=-q^2$.

By employing the aforementioned expressions, the hadronic description of the correlation function can be extracted as follows:
%
%
\begin{align}
\label{edmn09}
 \Pi_{\alpha\beta}^{Had}(p,q) &=  \frac{\varepsilon_\rho \, \lambda_{Y_{ c \bar c}}^2}{ [m_{Y_{ c \bar c}}^2 - (p+q)^2][m_{Y_{ c \bar c}}^2 - p^2]}
 \Bigg\{G_1(Q^2)(2p+q)_\rho\Bigg[g_{\alpha\beta}-\frac{p_\alpha p_\beta}{m_{Y_{ c \bar c}}^2}
 -\frac{(p+q)_\alpha (p+q)_\beta}{m_{Y_{ c \bar c}}^2}+\frac{(p+q)_\alpha p_\beta}{2m_{Y_{ c \bar c}}^4}\nonumber\\
 & \times (Q^2+2m_{Y_{ c \bar c}}^2)
 \Bigg]
 + G_2 (Q^2) \Bigg[q_\alpha g_{\rho\beta}  
 - q_\beta g_{\rho\alpha}-
\frac{p_\beta}{m_{Y_{ c \bar c}}^2}  \big(q_\alpha p_\rho - \frac{1}{2}
Q^2 g_{\alpha\rho}\big) 
+
\frac{(p+q)_\alpha}{m_{Y_{ c \bar c}}^2}  \big(q_\beta (p+q)_\rho+ \frac{1}{2}
Q^2 g_{\beta\rho}\big) 
\nonumber\\
&-  
\frac{(p+q)_\alpha p_\beta p_\rho}{m_{Y_{ c \bar c}}^4} \, Q^2
\Bigg]
-\frac{G_3(Q^2)}{m_{Y_{ c \bar c}}^2}(2p+q)_\rho \Bigg[
q_\alpha q_\beta -\frac{p_\alpha q_\beta}{2 m_{Y_{ c \bar c}}^2} Q^2 
+\frac{(p+q)_\alpha q_\beta}{2 m_{Y_{ c \bar c}}^2} Q^2
-\frac{(p+q)_\alpha q_\beta}{4 m_{Y_{ c \bar c}}^4} Q^4\Bigg]
\Bigg\}\,.
\end{align}

In a typical scenario, the aforementioned equation would suffice. However, given that our analysis pertains to real photons, it is necessary to derive expressions at the point $Q^2 = 0$, which represents the static limit. At this juncture, the aforementioned form factors are expressed in terms of the magnetic moment ($\mu$), resulting in the following expression:
\begin{align}
\label{edmn08}
 \mu  &= \frac{e}{2 m_{Y_{ c \bar c}}}\,G_2(0). 
\end{align}

The requisite formula for the magnetic moment is established, thus facilitating the formulation of the hadronic description of the calculation. The following step is to determine the correlation function in regard to the characteristics associated with quark-gluon interactions.

 \subsubsection{QCD Description}
 
In the context of the QCD description of the correlation function, the relevant interpolating currents for the identified vector hidden-charm tetraquark states are embedded within the correlation function expressed in Eq. (\ref{edmn01}). Subsequently, Wick's theorem is employed to execute all pertinent contractions and derive the relevant expressions. As a result of the above-mentioned procedure, the QCD description of the correlation function is extracted as follows:
%
\begin{align}
\label{eq:QCDSide}
\Pi _{\alpha \beta }^{\mathrm{QCD},\,J_\alpha^1}(p,q)&=i\varepsilon^{abc}\varepsilon^{a^{\prime}b^{\prime}c^{\prime}}\varepsilon^{ade}
\varepsilon^{a^{\prime}d^{\prime}e^{\prime}}\int d^{4}xe^{ip\cdot x}   \langle 0 \mid  
\Big[ \gamma _{5}{S}_{c}^{cc^{\prime }}(x)\gamma _{5} \widetilde S_{q_1}^{bb^{\prime }}(x)\Big]    
\mathrm{Tr}\Big[ \gamma _{5} \gamma _{\alpha } \widetilde{S}_{q_2}^{cc^{\prime}}(-x)\gamma _{\beta } \gamma _{5} S_{c}^{d^{\prime }d}(-x)\Big]   \mid 0 \rangle_{F} ,  \\
%
\Pi _{\alpha \beta }^{\mathrm{QCD},\,J_\alpha^2}(p,q)&=i\varepsilon^{abc}\varepsilon^{a^{\prime}b^{\prime}c^{\prime}}\varepsilon^{ade}
\varepsilon^{a^{\prime}d^{\prime}e^{\prime}}\int d^{4}xe^{ip\cdot x}   \langle 0 \mid  
\Big[ \gamma _{\alpha }\gamma _{5}{S}_{c}^{cc^{\prime }}(x)\gamma _{5} \gamma _{\beta } \widetilde S_{q_1}^{bb^{\prime }}(x)\Big]    
\mathrm{Tr}\Big[ \gamma _{5}  \widetilde{S}_{q_2}^{cc^{\prime}}(-x) \gamma _{5} S_{c}^{d^{\prime }d}(-x)\Big]   \mid 0 \rangle_{F} , 
\end{align}
\begin{align}
\Pi _{\alpha \beta }^{\mathrm{QCD},\,J_\alpha^3}(p,q)&=i\varepsilon^{abc}\varepsilon^{a^{\prime}b^{\prime}c^{\prime}}\varepsilon^{ade}
\varepsilon^{a^{\prime}d^{\prime}e^{\prime}}\int d^{4}xe^{ip\cdot x}   \langle 0 \mid  
\Big[ {S}_{c}^{cc^{\prime }}(x) \widetilde S_{q_1}^{bb^{\prime }}(x)\Big]    
\mathrm{Tr}\Big[  \gamma _{\alpha } \widetilde{S}_{q_2}^{cc^{\prime}}(-x)\gamma _{\beta }  S_{c}^{d^{\prime }d}(-x)\Big]   \mid 0 \rangle_{F} ,\\
%
\Pi _{\alpha \beta }^{\mathrm{QCD},\,J_\alpha^4}(p,q)&=i\varepsilon^{abc}\varepsilon^{a^{\prime}b^{\prime}c^{\prime}}\varepsilon^{ade}
\varepsilon^{a^{\prime}d^{\prime}e^{\prime}}\int d^{4}xe^{ip\cdot x}   \langle 0 \mid  
\Big[ \gamma _{\alpha }{S}_{c}^{cc^{\prime }}(x) \gamma _{\beta } \widetilde S_{q_1}^{bb^{\prime }}(x)\Big]    
\mathrm{Tr}\Big[  \widetilde{S}_{q_2}^{cc^{\prime}}(-x)  S_{c}^{d^{\prime }d}(-x)\Big]   \mid 0 \rangle_{F} ,  \label{eq:QCDSide3}
\end{align}
where $\widetilde{S}_{c(q)}^{ij}(x)=CS_{c(q)}^{ij\rm{T}}(x)C$. 
The corresponding propagators for heavy and light quarks are written as follows~\cite{Yang:1993bp, Belyaev:1985wza}:
\begin{align}
\label{edmn13}
S_{q}(x)&= S_q^{free}(x) 
- \frac{\langle \bar qq \rangle }{12} \Big(1-i\frac{m_{q} \xslash}{4}   \Big)
- \frac{ \langle \bar qq \rangle }{192}
m_0^2 x^2  \Big(1 
  -i\frac{m_{q} \xslash}{6}   \Big)
+\frac {i g_s~G^{\alpha \beta} (x)}{32 \pi^2 x^2} 
\bigg[\rlap/{x} 
\sigma_{\alpha \beta} +  \sigma_{\alpha \beta} \rlap/{x}
 \bigg],\\
%
S_{c}(x)&=S_c^{free}(x)
-\frac{m_{c}\,g_{s}\, G^{\alpha \beta}(x)}{32\pi ^{2}} \bigg[ (\sigma _{\alpha \beta }{\xslash}
+{\xslash}\sigma _{\alpha \beta }) 
    \frac{K_{1}\big( m_{c}\sqrt{-x^{2}}\big) }{\sqrt{-x^{2}}}
 +2\sigma_{\alpha \beta }K_{0}\big( m_{c}\sqrt{-x^{2}}\big)\bigg],
 \label{edmn14}
\end{align}%
with  
\begin{align}
 S_q^{free}(x)&=\frac{1}{2 \pi x^2}\Big(i \frac{\xslash}{x^2}- \frac{m_q}{2}\Big),\\
 S_c^{free}(x)&=\frac{m_{c}^{2}}{4 \pi^{2}} \bigg[ \frac{K_{1}\big(m_{c}\sqrt{-x^{2}}\big) }{\sqrt{-x^{2}}}
+i\frac{{\xslash}~K_{2}\big( m_{c}\sqrt{-x^{2}}\big)}
{(\sqrt{-x^{2}})^{2}}\bigg],
\end{align}
where $\langle \bar qq \rangle $ being light quark condensate, $G^{\alpha\beta}$ is the gluon field-strength tensor,  $m_0$ being the quark-gluon mixed condensate with $ m_0^2= \langle 0 \mid \bar  q\, g_s\, \sigma_{\alpha\beta}\, G^{\alpha\beta}\, q \mid 0 \rangle / \langle \bar qq \rangle $,  and $K_i$'s are the Bessel functions.  

The analytical expressions in Eqs. (\ref{eq:QCDSide})-(\ref{eq:QCDSide3}) are subject to contributions from both perturbative and non-perturbative interactions. The perturbative contributions arise from the photon interacting with light and heavy quark propagators at a short distance, whereas the non-perturbative contributions result from the photon interacting with light quarks at a large distance. In order to ensure the consistency and reliability of the analysis, it is essential to include these contributions. The aforementioned contributions are incorporated into the analysis via the following scheme: 
\begin{itemize}
 \item In the event that perturbative contributions are to be included in the analysis, it is sufficient to employ the following formulation. 
\begin{align}
\label{free}
S_{c(q)}^{free}(x) \longrightarrow \int d^4z\, S_{c(q)}^{free} (x-z)\,\rlap/{\!A}(z)\, S_{c(q)}^{free} (z)\,.
\end{align}
In this formulation, one of the quark propagators is replaced with a substitute in the equation above, while only the free part of the remaining three propagators is utilized.  This amounts to taking $\bar T_4^{\gamma} (\underline{\alpha}) = 0$ and $S_{\gamma} (\underline {\alpha}) = \delta(\alpha_{\bar q})\delta(\alpha_{q})$ as the light-cone distribution amplitude in the three particle distribution amplitudes (see Ref. \cite{Li:2020rcg}).  

\item In order to take account of non-perturbative elements in the course of the present analysis, it is recommended that the relevant formula be applied as follows:
 \begin{align}
\label{edmn21}
S_{q,\alpha\beta}^{ab}(x) \longrightarrow -\frac{1}{4} \big[\bar{q}^a(x) \Gamma_i q^b(0)\big]\big(\Gamma_i\big)_{\alpha\beta},
\end{align}
where $\Gamma_i = \{\textbf{1}$, $\gamma_5$, $\gamma_\alpha$, $i\gamma_5 \gamma_\alpha$, $\sigma_{\alpha\beta}/2\}$.  In this formulation, one of the quark propagators is replaced with a substitute in the equation above, while the remaining three propagators are considered full.

The application of non-perturbative components to the analytical process results in the appearance of matrix elements, such as $\langle \gamma(q)\vel \bar{q}(x) \Gamma_i G_{\alpha\beta}q(0) \ver 0\rangle$ and $\langle \gamma(q)\vel \bar{q}(x) \Gamma_i q(0) \ver 0\rangle$. The aforementioned matrix elements are defined by the photon distribution amplitudes (DAs)~\cite{Ball:2002ps}. 
A detailed overview of the methodologies utilized to integrate both perturbative and non-perturbative elements into the computations can be elucidated in Refs.~\cite{Ozdem:2022vip, Ozdem:2022eds}. After implementing the previously mentioned modifications, which entail considering both perturbative and non-perturbative contributions in the computations, the QCD description of the correlation function is derived. 
\end{itemize}

\subsubsection{QCD light-cone sum rules for magnetic moments}

In conclusion, by matching the corresponding coefficients of the distinctive Lorentz structure ($q_\alpha \varepsilon_\beta - \varepsilon_\alpha q_\beta$) from the QCD and hadronic descriptions, we obtain the QCD light-cone sum rules, which allow us to determine the responsible magnetic moments in terms of QCD and hadronic parameters, as well as auxiliary parameters $\rm{s_0}$ and $\rm{M^2}$. The results obtained are as follows: 
\begin{align}
\label{jmu1}
 \mu_{Y_{ c \bar c}}^{J_\alpha^1}\,\lambda^{2, {J_\alpha^1}}_{Y_{ c \bar c}}  &= e^{\frac{m_{Y_{ c \bar c}}^{2,{J_\alpha^1}}}{\rm{M^2}}} \,\, \rho_1(\rm{M^2},\rm{s_0}),~~~~
 \mu_{Y_{ c \bar c}}^{J_\alpha^2}\, \lambda^{2, {J_\alpha^2}}_{Y_{ c \bar c}}  = e^{\frac{m_{Y_{ c \bar c}}^{2, {J_\alpha^2}}}{\rm{M^2}}} \,\, \rho_2(\rm{M^2},\rm{s_0}),\\
  \nonumber\\
  \mu_{Y_{ c \bar c}}^{J_\alpha^3}\,  \lambda^{2,{J_\alpha^3}}_{Y_{ c \bar c}} &= e^{\frac{m_{Y_{ c \bar c}}^{2, {J_\alpha^3}}}{\rm{M^2}}} \,\, \rho_3(\rm{M^2},\rm{s_0}),~~~~
 \mu_{Y_{ c \bar c}}^{J_\alpha^4}\, \lambda^{2,{J_\alpha^4}}_{Y_{ c \bar c}}  = e^{\frac{m_{Y_{ c \bar c}}^{2,{J_\alpha^4}}}{\rm{M^2}}} \,\, \rho_4(\rm{M^2},\rm{s_0}).\label{jmu4}
 \end{align}

The expressions for the $\rho_i(\rm{M^2},\rm{s_0})$ functions are presented in the following manner,
\begin{align}
 \rho_1(\rm{M^2},\rm{s_0})&= \frac{1}{2 ^{20} \times 3^2 \times 5^2 \times 7 \pi^5}
 \Big[  14 e_c \Big (6 I[0, 6] - 
    2 m_c \big ((19 m_ {q_ 1} - 72 m_ {q_ 2}) I[0, 5] + 
       65 m_ {q_ 1} I[1, 4]\big) + 27 I[1, 5]\Big) \nonumber\\
       &- 
 9 e_ {q_ 2} \Big (4 I[0, 6] + 
    14  m_ {q_ 1} m_c (I[0, 5] + 35 I[1, 4]) + 129 I[1, 5]\Big) \Big]
    \nonumber\\
    &+\frac{ e_{q_1} m_c \chi \langle g_s^2 G^2\rangle \langle \bar q_1 q_1 \rangle }{2 ^{17} \times 3^2 \times 5 \pi^3} I_5[\varphi_\gamma] I[0, 3]
    \nonumber\\
    &+\frac{ m_c \langle \bar q_1 q_1 \rangle }{2 ^{18} \times 3 \times 5 \pi^3} \Big[
      45 e_{q_1} I_2[\mathcal S] I[0, 4] \Big]
    \nonumber\\
    &+\frac{ m_c \langle \bar q_2 q_2 \rangle }{2 ^{14} \times 3^3 \times 5^2 \pi^3} \Big[
     80 e_ {q_2}  \big (44 m_{q_1} m_c I[0, 3] - 9 I[0, 4] \big) I_6[h_\gamma]  + 
9 \chi e_{q_2}  \big (20 m_{q_1} m_c I[0, 4] - 
    7 I[0, 5]\big) I_6[\varphi_\gamma]\Big]
    \nonumber\\
    &+\frac{f_{3\gamma}}{2 ^{23} \times 3 \times 5^2 \times 7\pi^3}
    \Big[ 153 e_{q_1} I_2[\mathcal V] I[0, 5] - 
 896 e_{q_2} I_5[\psi^a] \big(35 m_{q_1} m_c I[0, 4] + 9 I[0, 5]\big) \Big],\\
 \rho_2(\rm{M^2},\rm{s_0})&= -\frac{1}{2 ^{20} \times 3^2 \times 5^2 \times 7 \pi^5}\Big[  14 m_c \big(144 e_c m_{q_1} + 9 e_{q_1} m_{q_2} - 38 e_c m_{q_2}\big)  I[0, 5] + 
 12 (3 e_{q_1} + 7 e_c) I[0, 6] \nonumber\\
    &+ 
 70 m_{q_2} m_c(63 e_{q_1} - 26 e_c)  I[1, 4] + 27 (43 e_{q_1} + 14 e_c) I[1, 5]  \Big]
    \nonumber\\
    &+\frac{ e_{q_2} m_c \chi \langle g_s^2 G^2\rangle \langle \bar q_2q_2 \rangle }{2 ^{18} \times 3^2 \times 5 \pi^3} I_5[\varphi_\gamma] I[0, 3]
    \nonumber\\
    &+\frac{e_{q_1} m_c \langle \bar q_1 q_1 \rangle }{2 ^{15} \times 3^3 \times 5^2 \pi^3} \Big[
    -20 m_c   m_ {q_ 2} \Big (15 I_ 3[\mathcal S] - 
        15 I_ 3[\mathcal {\tilde S}] + 22 I_ 6[h_\gamma]\Big) I[
   0, 3] + 45 \Big ( 
     I_ 3[\mathcal S] - I_ 3[\mathcal {\tilde S}] \nonumber\\
    & + 
        2 I_ 6[h_\gamma] - 
        8 \chi m_ {q_ 2} m_c I_ 6[\varphi_\gamma]\Big) I[0, 4] + 
 126 \chi  I_ 6[\varphi_\gamma] I[0, 5]
    \Big]
    \nonumber\\
    &+\frac{e_{q_2} m_c \langle \bar q_2 q_2 \rangle }{2 ^{18} \times 3^3 \times 5 \pi^3} \Big[
   (240 m_{q_1} m_c I[0, 3] + 22 I[0, 4])I_4[\mathcal S] -17 I_1[\mathcal S] I[0, 4]  
    \Big]
    \nonumber\\
    &+\frac{f_{3\gamma}}{2 ^{23} \times 3^3 \times 5^2 \times 7\pi^3}
    \Big[ 2240  m_c \Big (9 e_ {q_ 2} m_ {q_ 1} I_ 1[\mathcal V] + 
    e_ {q_ 1} m_ {q_ 2}\big (-11 I_ 2[\mathcal V] + 
        126 I_ 5[\psi^a]\big)\Big) I[0, 4] 
        \nonumber\\
    &+ 
 9  \Big (279 e_ {q_ 2} I_ 1[\mathcal V] + 
    448 e_ {q_ 1}\big (I_ 2[\mathcal V] + 18 I_ 5[\psi^a])\Big) I[0, 
   5] \Big],\\
%
 \rho_3(\rm{M^2},\rm{s_0})&= \frac{1}{2 ^{20} \times 3^2 \times 5^2 \times 7 \pi^5}
 \Big[  9 e_ {q_ 2} \Big (-4 I[0, 6] + 
    14 m_ {q_ 1} m_c (I[0, 5] + 35 I[1, 4]) - 129 I[1, 5]\Big) + 
 14 e_c \Big (6 I[0, 6] \nonumber\\
    &+ 
    2 m_c (19 m_ {q_ 1} - 72 m_ {q_ 2}) I[0, 5] + 
    130 m_c m_ {q_ 1} I[1, 4] + 27 I[1, 5]\Big) \Big]
    \nonumber\\
    &-\frac{ e_{q_1} m_c \chi \langle g_s^2 G^2\rangle \langle \bar q_1 q_1 \rangle }{2 ^{17} \times 3^2 \times 5 \pi^3} I_5[\varphi_\gamma] I[0, 3]
    \nonumber\\
        &-\frac{e_ {q_ 1} m_c \langle \bar q_1 q_1 \rangle }{2 ^{18} \times 3 \times 5 \pi^3} \Big[
        I_ 2[\mathcal S] I[0, 4] \Big]
    \nonumber\\
    &+\frac{e_ {q_ 2} m_c \langle \bar q_2 q_2 \rangle }{2 ^{14} \times 3^3 \times 5^2 \pi^3} \Big[
   5(44 m_ {q_ 1} m_c I[0, 3] + 
    9 I[0, 4]) I_ 6[\varphi_\gamma] + 
 9 \chi   (20 m_ {q_ 1} m_c I[0, 4] + 
    7 I[0, 5]) I_ 6[\varphi_\gamma]\Big]
    \nonumber\\
    &+\frac{f_{3\gamma}}{2 ^{23} \times 3 \times 5^2 \times 7\pi^3}
    \Big[ 896 e_ {q_ 2}  I_ 5[\psi^a] \big (35 m_ {q_ 1} m_c I[0, 4] - 
    9 I[0, 5]\big) + 153 e_ {q_ 1} I_ 2[\mathcal V] I[0, 5]\big) \Big],
 \end{align}
   \begin{align}
 \rho_4(\rm{M^2},\rm{s_0})&= \frac{1}{2 ^{20} \times 3^2 \times 5^2 \times 7 \pi^5}\Big[  28 e_c \Big (m_c (72 m_ {q_ 1} - 19 m_ {q_ 2})  I[0, 5] - 3 I[0, 6] - 
    65 m_ {q_ 2} m_c I[1, 4]\Big) + 
 18 e_ {q_ 1} \Big (-2 I[0, 6] \nonumber\\
    &+ 
    7 m_ {q_ 2} m_c \big (I[0, 5] + 35 I[1, 4]\big)\Big) - 
 27 (43 e_ {q_ 1} + 14 e_c) I[1, 5] \Big]
%
    \nonumber\\
    &-\frac{ e_{q_2} m_c \chi \langle g_s^2 G^2\rangle \langle \bar q_2 q_2 \rangle }{2 ^{18} \times 3^2 \times 5 \pi^3} I_5[\varphi_\gamma] I[0, 3]
    \nonumber\\
    &+\frac{ e_ {q_ 1} m_c \langle \bar q_1 q_1 \rangle }{2 ^{15} \times 3^3 \times 5^2 \pi^3} \Big[
     -20 m_ {q_ 2}m_c \Big ( - 
    15 I_ 3[\mathcal S] - 
        15 I_ 3[\mathcal {\tilde S}] + 22 I_ 6[h_\gamma]\Big) I[
   0, 3] - 45 \Big (    I_ 3[\mathcal S] - I_ 3[\mathcal {\tilde S}] + 
        2 I_ 6[h_\gamma]\nonumber\\
    & + 
        8 \chi m_ {q_ 2} m_c I_ 6[h_\gamma]\Big) I[0, 4] 
        - 
 126 \chi e_ {q_ 1} I_ 6[\varphi_\gamma] I[0, 5]  \Big]
    \nonumber\\
    &+\frac{ e_ {q_ 2} m_c \langle \bar q_2 q_2 \rangle }{2 ^{18} \times 3^3 \times 5 \pi^3} \Big[ (240 m_ {q_ 1} m_c I[0, 3] - 22 I[0, 4]) I_ 4[\mathcal S] + 
 17 I_ 1[\mathcal S] I[0, 4]  \Big]
    \nonumber\\
    &+\frac{f_{3\gamma}}{2 ^{23} \times 3^3 \times 5^2 \times 7\pi^3}
    \Big[ -2240  m_c \Big (9 e_ {q_ 2} m_ {q_ 1} I_ 1[\mathcal V] + 
    e_ {q_ 1} m_ {q_ 2} \big (-11 I_ 2[\mathcal V] + 
        126 I_ 5[\psi^a]\big)\Big) I[0, 4]\nonumber\\
    & + 
 9  \Big (279 e_ {q_ 2} I_ 1[\mathcal V] + 
    448 e_ {q_ 1} \big (I_ 2[\mathcal V] + 18 I_ 5[\psi^a])\Big) I[0, 
   5] \Big],
\end{align}

\noindent where $\langle g_s^2 G^2\rangle$ is gluon condensate;  $\langle \bar q_1 q_1 \rangle$ and $\langle \bar q_2 q_2 \rangle$ stand for corresponding light-quark condensates, respectively. It should be noted that the analytical expressions include terms such as $m_0^2 \langle \bar q_1 q_1 \rangle \langle \bar q_2 q_2 \rangle$, $m_0^2 \langle \bar q_1 q_1 \rangle f_{3\gamma}$, $m_0^2 \langle \bar q_2 q_2 \rangle f_{3\gamma}$ and $\langle \bar q_1 q_1 \rangle \langle \bar q_2 q_2 \rangle$. However, as the contribution of these terms is considerably small, they are not included here, although they are taken into account in the numerical analysis. The $I[n,m]$, and~$I_i[\mathcal{F}]$ functions are expressed as: 
\begin{align}
 I[n,m]&= \int_{\mathcal M}^{\rm{s_0}} ds ~ e^{-s/\rm{M^2}}~
 s^n\,(s-\mathcal M)^m,\nonumber\\
 I_1[\mathcal{F}]&=\int D_{\alpha_i} \int_0^1 dv~ \mathcal{F}(\alpha_{\bar q},\alpha_q,\alpha_g)
 \delta'(\alpha_ q +\bar v \alpha_g-u_0),\nonumber\\
  I_2[\mathcal{F}]&=\int D_{\alpha_i} \int_0^1 dv~ \mathcal{F}(\alpha_{\bar q},\alpha_q,\alpha_g)
 \delta'(\alpha_{\bar q}+ v \alpha_g-u_0),\nonumber\\
   I_3[\mathcal{F}]&=\int D_{\alpha_i} \int_0^1 dv~ \mathcal{F}(\alpha_{\bar q},\alpha_q,\alpha_g)
 \delta(\alpha_ q +\bar v \alpha_g-u_0),\nonumber\\
   I_4[\mathcal{F}]&=\int D_{\alpha_i} \int_0^1 dv~ \mathcal{F}(\alpha_{\bar q},\alpha_q,\alpha_g)
 \delta(\alpha_{\bar q}+ v \alpha_g-u_0),\nonumber\\
    I_5[\mathcal{F}]&=\int_0^1 du~ \mathcal{F}(u)\delta'(u-u_0),\nonumber\\
 I_6[\mathcal{F}]&=\int_0^1 du~ \mathcal{F}(u),\nonumber
 \end{align}
 where  $\mathcal M = 4m_c^2$ for $[u c][\bar c \bar d]$ states, $\mathcal M = (2m_c+m_s)^2$ for $[u c][\bar c \bar s]$ and $[d c][\bar c \bar s]$  states, and $\mathcal M = (2m_c+2m_s)^2$ for $[s c][\bar c \bar s]$  states; and $\mathcal{F}$ denotes the relevant DAs of the photon.

\end{widetext}

\section{Numerical illustrations}\label{numerical}

In order to determine magnetic moments through the evaluation of QCD light-cone sum rules, several input quantities are required. 
In conducting the magnetic moment analysis, the following values are employed: 
$m_s =93.4^{+8.6}_{-3.4}\,\mbox{MeV}$, $m_c = 1.27 \pm 0.02\,\mbox{GeV}$~\cite{ParticleDataGroup:2022pth}, $\langle \bar uu\rangle = 
\langle \bar dd\rangle=(-0.24 \pm 0.01)^3\,\mbox{GeV}^3$, $\langle \bar ss\rangle = 0.8\, \langle \bar uu\rangle$ $\,\mbox{GeV}^3$ \cite{Ioffe:2005ym},
$m_0^{2} = 0.8 \pm 0.1 \,\mbox{GeV}^2$ \cite{Ioffe:2005ym}, $\chi= -2.85 \pm 0.5 $ GeV$^{-2}$ \cite{Rohrwild:2007yt}, $f_{3\gamma}=-0.0039~$GeV$^2$~\cite{Ball:2002ps}, and $\langle g_s^2G^2\rangle = 0.48 \pm 0.14~ \mbox{GeV}^4$~\cite{Narison:2018nbv}. In numerical calculations, we set $m_u$ =$m_d$ = 0 and $m_s^2=0$, but consider terms proportional to $m_s$.   The mass and residue values for these states are essential for further analysis and have been taken from the Ref.~\cite{Wang:2009gx}. 
One of the important input parameters in our numerical calculations is photon DAs. The aforementioned DAs and input parameters, as utilized in their explicit forms, are presented in the Ref.~\cite{Ball:2002ps}.

As demonstrated by the expressions in Eqs. (\ref{jmu1})$-$(\ref{jmu4}), two auxiliary parameters, $\rm{s_0}$ and $\rm{M^2}$, have been identified. To ensure the reliability and consistency of our analyses, it is essential to determine the region, designated as the working region, within which the magnetic moment results should exhibit slight variation concerning these parameters. The working regions of these auxiliary parameters are defined by pole dominance (PC) and convergence of OPE (CVG), which represent the standard procedures of the methodology employed. The aforementioned procedures are formalized as follows:
\begin{align}
 \mbox{PC} &=\frac{\rho_i (\rm{M^2},\rm{s_0})}{\rho_i (\rm{M^2},\infty)} \geq 30 \%, ~~~~~~
\\
 \mbox{CVG} (\rm{M^2}, \rm{s_0}) &=\frac{\rho_i^{\rm{Dim N}} (\rm{M^2},\rm{s_0})}{\rho_i (\rm{M^2},\rm{s_0})}\leq 5\%,
 \end{align}
 where $\rho_i^{\rm{Dim N}} (\rm{M^2},\rm{s_0})$ is the highest dimensional term in the OPE  of $\rho_i (\rm{M^2},\rm{s_0})$. 
 Following the aforementioned criteria, the working regions of the auxiliary parameters, as delineated in Table \ref{table}, are determined. It can be easily seen from the results presented in the table that the requirements of the method are fulfilled. At this point in our investigation, we have met all of the criteria intrinsic to the method used and anticipate that our predictions will be reliable. To provide a more comprehensive analysis, Figs.~\ref{figMsq}-\ref{figMsq14} illustrate the variations in the derived magnetic moments of these states for auxiliary parameters. 
  As illustrated in the figures, the magnetic moments of these states show a relatively mild dependence on the aforementioned auxiliary variables.

The calculated magnetic moments of vector hidden-charm tetraquark states, which account for the uncertainty associated with the input variables and the variation in the auxiliary parameters, are presented in Table~\ref{table}.  Furthermore, to facilitate improved interpretation, we have included the outcomes of the magnetic moments, illustrated alongside both their central values and the combined values with associated errors, in Figs. \ref{figMsq1} and \ref{figMsq2}. 

In view of the results obtained in this study, the following comments are offered for consideration:
  \begin{itemize}
    
\item Upon examination of the results obtained, it is evident that the contribution of the short-distance interaction of quarks with the photon, i.e. the perturbative component, constitutes the dominant factor in the analysis, as can be seen in the Table \ref{table}. The remaining contributions are derived from the long-distance interactions of light quarks with the photon, which is a non-perturbative contribution.
 
\item Our analysis demonstrates that the magnetic moments of these states are determined by the vector diquark ($[\bar c \bar q_2]$ component for $J_\alpha^1$ and $J_\alpha^3$, and $[\bar c \bar q_1]$ for $J_\alpha^2$ and $J_\alpha^4$) component of the interpolating currents. The results may reveal a direct correlation between the electromagnetic characteristics and the internal configuration of the examined states.
  
\item The individual quark contributions to the magnetic moment results have also been analyzed and the results are presented in Table \ref{table3}. It should be noted that the central values of the input parameters are employed in these analyses. A detailed investigation into the individual quark contributions to the magnetic moments reveals that the results are considerably influenced by light quarks, with the majority of the contribution coming from these quarks. The contribution of the c-quark to the magnetic moment results is approximately in the range of ($23-47)\%$, while the contribution of light quarks varies in the range of ($53-77)\%$. As can be seen from this analysis, the individual quark contributions in the results vary depending on the diquark-antidiquark structure chosen.

\item The ordering of the magnetic moments could provide insight into the experimental accessibility of corresponding physical quantities. Given the magnitude of their ordering, these outcomes may be attainable in forthcoming experiments. 
  
\item As a byproduct, the quadrupole moments ($\mathcal D$) of these states are also extracted, as illustrated in Table \ref{table2}. It can be seen that the results are small compared to the magnetic moment results. As with the magnetic moment results, the quadrupole moment results vary according to the nature of the interpolating current selected. The quadrupole moment results obtained for these states are non-zero, indicating the presence of a non-spherical charge distribution. It is well known that the geometrical shape of the hadrons under investigation can be interpreted by means of the sign of the quadrupole moments. If the predicted sign is negative, the shape of the hadron is considered to be oblate; if it is positive, it is considered to be prolate.  Upon analysis of the quadrupole moment results, it can be observed that, with the exception of the $[dc][\bar c \bar s]$ and $[sc][\bar c \bar s]$ tetraquark states obtained through the use of $J_\alpha^2 $ and $J_\alpha^4 $ interpolating currents, the results of the remaining states exhibited a negative sign and could be characterized as oblate in shape. 

\item It is indeed possible to determine the ratio of $U$-symmetry violation in the obtained magnetic and quadrupole moment results $\Big(\frac{[uc][\bar c \bar d]}{[uc][\bar c \bar s]}\Big)$ and $ \Big(\frac{[dc][\bar c \bar s]}{[sc][\bar c \bar s]}\Big)$.  The $U$-symmetry violation has been contemplated with the consideration of a non-zero s-quark mass and disparate s-quark condensates.The observed deviation from the $U$-symmetry violation in the magnetic and quadrupole moment results reaches a maximum of approximately $20\%$, which appears to be a reasonable value.

\item As a final and key comment, as illustrated in Table~\ref{table2}, the different interpolating currents used to examine the electromagnetic characteristics of the vector hidden-charm tetraquark states, which are constituted of the same quarks, result in notable discrepancies in the outcomes derived from their application.  It is plausible that the aforementioned phenomenon could be interpreted as an indication of the existence of more than one vector hidden-charm tetraquark states, characterized by identical quantum numbers and a similar composition of quarks, yet differing in their electromagnetic characteristics. As previously stated, the interpolating currents in question possess identical quantum numbers, thereby giving rise to nearly degenerate masses for the aforementioned vector hidden-charm tetraquark states~\cite{Wang:2009gx}. Nevertheless, the results obtained for electromagnetic characteristics are highly responsive to the diquark-antidiquark configuration and the intrinsic attributes of the state under investigation.  It is typically presumed that modifying the basis of hadrons will not influence the resulting data. However, it cannot be ruled out that this assumption may not be applicable in the context of electromagnetic characteristics.  This is due to the fact that the electromagnetic characteristics of the aforementioned states are directly correlated with their internal structural organization. In the terminology of electromagnetic characteristics, a modification to the basis for the associated state also entails a transformation in the internal structure of the hadron; this, in turn, may give rise to a notable alteration in the computed outcomes. In Refs.~\cite{Ozdem:2024dbq, Azizi:2023gzv, Ozdem:2024rqx, Ozdem:2022iqk, Ozdem:2024rch}, several different interpolating currents have been used in order to obtain the electromagnetic characteristics of both tetra- and pentaquarks. The studies yielded significant discrepancies in the magnetic moments obtained when using diquark-antidiquark or diquark-diquark-antiquark structures.    Accordingly, the selection of distinct interpolating currents capable of coupling with the same states, or alternatively the alteration of the isospin and charge basis of the states under examination, may consequently lead to disparate magnetic moments. 

\end{itemize}

\section{Summary and outlook}\label{sum}
Accessing a full picture of the internal structure of hadrons would be a key topic of hadron physics, with the main motivation to study the strong interaction binding the visible matter.  Furthermore, the underlying structure of known exotic states remains an unresolved fundamental issue in hadron physics, which is currently being addressed by hadron physics community. 
The magnetic moment provides an ideal platform for investigating the internal structure of particles as governed by the quark-gluon dynamics of QCD. This is due to its status as the primary response of a bound system to a weak external magnetic field. In light of this, this work  aims to determine the magnetic moments of vector hidden-charm tetraquark states by making use of QCD light-cone sum rules. In order to achieve this objective, the states mentioned above are considered in terms of the diquark-antidiquark structure. Subsequently, a comprehensive examination is conducted, with four distinct interpolating currents being given particular consideration, as these have the potential to couple with the aforementioned states.   It has been observed that there are considerable discrepancies between the magnetic moment results extracted employing different diquark-antidiquark structures. Such a prediction may be interpreted as the possibility of more than one tetraquark with the identical quantum numbers and similar quark constituents, but with different magnetic moments. The numerical predictions yielded have led to the conclusion that the magnetic moments of the vector hidden-charm tetraquark states are capable of projecting the inner structure of these states, which may then be used to determine their quark-gluon structure and quantum numbers. In order to provide a comprehensive analysis, the individual quark contributions to the magnetic moments are also examined. As a byproduct, the quadrupole moments of these states are also extracted, and the obtained values differ from zero, which indicates that the charge distribution of these states is not spherical. It is our hope that our projections regarding the magnetic moments of vector hidden-charm tetraquark states, when considered alongside the findings of other theoretical investigations into these tetraquarks' spectroscopic attributes and decay widths, will prove advantageous in the pursuit of these states in forthcoming experiments and in elucidating the nature of these hadrons.

\begin{widetext}

  \begin{table}[htb!]
	\addtolength{\tabcolsep}{6pt}
	\caption{Working regions of $\rm{M^2}$ and $\rm{s_0}$
together with the PC and CVG for the magnetic moments of the vector hidden-charm tetraquark states, where the "perturbative" 
denotes the contributions from the perturbative terms.}
	\label{table}
	\begin{ruledtabular}
\begin{tabular}{lccccccc}
	   \\
	   Currents&Tetraquarks & $\rm{s_0}\,\,[\rm{GeV}^2]$&   $\rm{M^2}\,\,[\rm{GeV}^2]$& PC\,\,[$\%$] & CVG\,\,[$\%$]	& Perturbative\,\,[$\%$]    \\
	   \\
	   \hline\hline
&$[uc]~ [\bar c \bar d]$& [28.0, 30.0]&  [3.2, 3.8]& [64.39, 46.59]& $ \ll1$ & (83.0 $-$ 90.0)\\
~~~~$J_\alpha^1 $&$[uc] ~[\bar c \bar s]$& [29.0, 31.0]&  [3.3, 4.1]& [65.13, 42.84]& $\ll 1$& (83.0 $-$ 90.0)\\
&$[dc] ~[\bar c \bar s]$& [29.0, 31.0]& [3.3, 4.1]& [65.12, 42.92]& $ \ll1$& (82.0 $-$ 90.0)\\
&$[sc] ~[\bar c \bar s]$& [30.0, 32.0] & [3.3, 4.1]& [68.92, 46.52]& $\ll 1$& (83.5 $-$ 89.0)\\
\hline\hline
&$[uc]~ [\bar c \bar d]$& [28.0, 30.0]&  [3.2, 3.8]& [62.35, 44.48]& $\ll 1$ & (84.0 $-$ 87.0)\\
~~~~$J_\alpha^2 $&$[uc] ~[\bar c \bar s]$& [29.0, 31.0]&  [3.3, 4.1]& [63.45, 42.22]& $\ll 1$& (83.0 $-$ 88.0)\\
&$[dc] ~[\bar c \bar s]$& [29.0, 31.0]& [3.3, 4.1]& [64.01, 41.75]& $ \ll 1$ & (56.0 $-$ 67.0)\\
&$[sc] ~[\bar c \bar s]$ & [30.0, 32.0]& [3.3, 4.1]& [69.51, 46.36]& $ \ll 1$& (60.0 $-$ 70.0)\\
	   \hline\hline
&$[uc] ~[\bar c \bar d]$& [28.0, 30.0]&  [3.2, 3.8]& [62.57, 44.77]& $\ll 1$ &(84.0 $-$ 90.0)\\
~~~~$J_\alpha^3 $&$[uc] ~[\bar c \bar s]$& [29.0, 31.0]&  [3.3, 4.1]& [63.50, 41.49]& $\ll 1$ &(85.0 $-$ 91.0)\\
&$[dc]~ [\bar c \bar s]$& [29.0, 31.0]& [3.3, 4.1]& [63.64, 41.40]& $\ll 1$ &(85.0 $-$ 88.0)\\
&$[sc]~ [\bar c \bar s]$& [30.0, 32.0] & [3.3, 4.1]& [67.39, 45.20]& $\ll 1$ &(84.0 $-$ 87.0)\\
	   \hline\hline
&$[uc] ~[\bar c \bar d]$& [28.0, 30.0]&  [3.2, 3.8]& [64.59, 46.94]& $\ll 1$& (84.0 $-$ 87.0)\\
~~~~$J_\alpha^4 $&$[uc] ~[\bar c \bar s]$& [29.0, 31.0]&  [3.3, 4.1]& [65.42, 43.27]& $\ll 1$& (83.0 $-$ 88.0)\\
&$[dc] ~[\bar c \bar s]$& [29.0, 31.0]& [3.3, 4.1]& [66.66, 45.32]& $\ll 1$& (56.0 $-$ 68.0)\\
&$[sc] ~[\bar c \bar s]$ & [30.0, 32.0] & [3.3, 4.1]& [70.27, 48.36]& $\ll 1$& (62.0 $-$ 71.0)\\
\end{tabular}
\end{ruledtabular}
\end{table}

 \end{widetext}

\begin{widetext}
 
  \begin{table}[htb!]
	\addtolength{\tabcolsep}{6pt}
	\caption{The predicted magnetic moments of vector hidden-charm tetraquark states.}
	\label{table2}
	\begin{ruledtabular}
\begin{tabular}{lccccccc}
	   \\
	   Currents&Tetraquarks &   $\mu \,[\mu_N]$& $\mathcal{D}\, [\rm{fm}^2]\, (\times 10^{-2})$\\
	   \\
	   \hline\hline
&$[uc]~ [\bar c \bar d]$&  $~~3.75 \pm 0.50$& $-1.18 \pm 0.17$\\
~~~~$J_\alpha^1 $&$[uc] ~[\bar c \bar s]$&  $~~4.52 \pm 0.70$& $-1.41 \pm 0.25$\\
&$[dc] ~[\bar c \bar s]$& $~~4.51 \pm 0.70$& $-1.42 \pm 0.25$\\
&$[sc] ~[\bar c \bar s]$ & $~~4.85 \pm 0.66$& $-1.51 \pm 0.26$\\
\hline\hline
&$[uc]~ [\bar c \bar d]$&  $-4.68 \pm 0.62$& $-2.44 \pm 0.35$\\
~~~~$J_\alpha^2 $&$[uc]~ [\bar c \bar s]$&  $-5.83 \pm 0.90$& $-2.99 \pm 0.55$\\
&$[dc]~ [\bar c \bar s]$& $~~0.42 \pm 0.07$& $~~1.49 \pm 0.27$\\
&$[sc]~ [\bar c \bar s]$ & $~~0.50 \pm 0.07$& $~~1.55 \pm 0.27$\\
	   \hline\hline
&$[uc] ~[\bar c \bar d]$&  $~~2.67 \pm 0.34$& $-1.02 \pm 0.14$\\
~~~~$J_\alpha^3 $&$[uc] ~[\bar c \bar s]$&  $~~2.68 \pm 0.33$& $-0.95 \pm 0.15$\\
&$[dc]~ [\bar c \bar s]$& $~~2.68 \pm 0.33$& $-0.95 \pm 0.15$\\
&$[sc]~ [\bar c \bar s]$ & $~~3.34 \pm 0.38$& $-1.14 \pm 0.17$\\
	   \hline\hline
&$[uc]~ [\bar c \bar d]$&  $-5.26 \pm 0.67$& $-2.10 \pm 0.29$\\
~~~~$J_\alpha^4 $&$[uc] ~[\bar c \bar s]$&  $-5.15 \pm 0.65$& $-1.92 \pm 0.29$\\
&$[dc]~ [\bar c \bar s]$& $~~0.59 \pm 0.07$& $~~0.96 \pm 0.14$\\
&$[sc]~ [\bar c \bar s]$ & $~~0.70 \pm 0.08$& $~~1.18 \pm 0.18$\\
\end{tabular}
\end{ruledtabular}
\end{table}
 
 \end{widetext}
 
\begin{widetext}

  \begin{table}[htb!]
	\addtolength{\tabcolsep}{6pt}
	\caption{The individual quark contributions to the magnetic moments of vector hidden-charm tetraquark states.}
	\label{table3}
	\begin{ruledtabular}
\begin{tabular}{lccccccc}
	   \\
	   Currents&Tetraquarks &   $\mu_{q_1} \,[\mu_N]$& $\mu_c \,[\mu_N]$& $\mu_{q_2} \,[\mu_N]$&$\mu_{tot} \,[\mu_N]$\\
	  \\
	   \hline\hline
&$[uc] ~[\bar c \bar d]$&  $~~0.0034$& $~~1.448$& 2.298 &~~3.75\\
~~~~$J_\alpha^1 $&$[uc]~ [\bar c \bar s]$&  $~~0.0040$& $~~1.758$& 2.758&~~4.52\\
&$[dc]~ [\bar c \bar s]$& $-0.0020$& $~~1.839$& 2.673&~~4.51\\
&$[sc]~ [\bar c \bar s]$ & $-0.0021$& $~~1.895$ & 2.957&~~4.85\\
\hline\hline
&$[uc]~ [\bar c \bar d]$&  $-3.215$& $-1.462$& $-0.0030$& $-4.68$\\
~~~~$J_\alpha^2 $&$[uc]~ [\bar c \bar s]$&  $-3.926$& $-1.900$& $-0.0036$ &$-5.83 $\\
&$[dc] ~[\bar c \bar s]$& $~~1.963$& $-1.539$& $-0.0040$ &~~0.42\\
&$[sc] ~[\bar c \bar s]$ & $~~2.210$& $-1.706$ & $-0.0040$ & ~~0.50\\
	   \hline\hline
&$[uc]~ [\bar c \bar d]$&  $~~0.0032$& $~~1.260$& ~~1.406& ~~2.67\\
~~~~$J_\alpha^3 $&$[uc]~ [\bar c \bar s]$&  $~~0.0030$& ~~1.247& ~~1.431  & ~~2.68\\
&$[dc]~ [\bar c \bar s]$& $-0.0015$& $~~1.247$& ~~1.435 & ~~2.68\\
&$[sc]~ [\bar c \bar s]$ & $-0.0018$& $~~1.564$& ~~1.794 &~~3.34\\
	   \hline\hline
&$[uc] ~[\bar c \bar d]$&  $-4.019$& $-1.238$&$-0.003$ &$-5.26$\\
~~~~$J_\alpha^4 $&$[uc]~ [\bar c \bar s]$&  $-3.963$& $-1.247$& $-0.005$ & $-5.15$\\
&$[dc]~ [\bar c \bar s]$& $~~1.976$& $-1.383$& $-0.003$& ~~0.59\\
&$[sc] ~[\bar c \bar s]$ & $~~2.397$& $-1.693$& $-0.003$&~~0.70\\
\end{tabular}
\end{ruledtabular}
\end{table}
 
\end{widetext}

   \begin{widetext}
   
  \begin{figure}[htp]
\centering
  \subfloat[]{\includegraphics[width=0.47\textwidth]{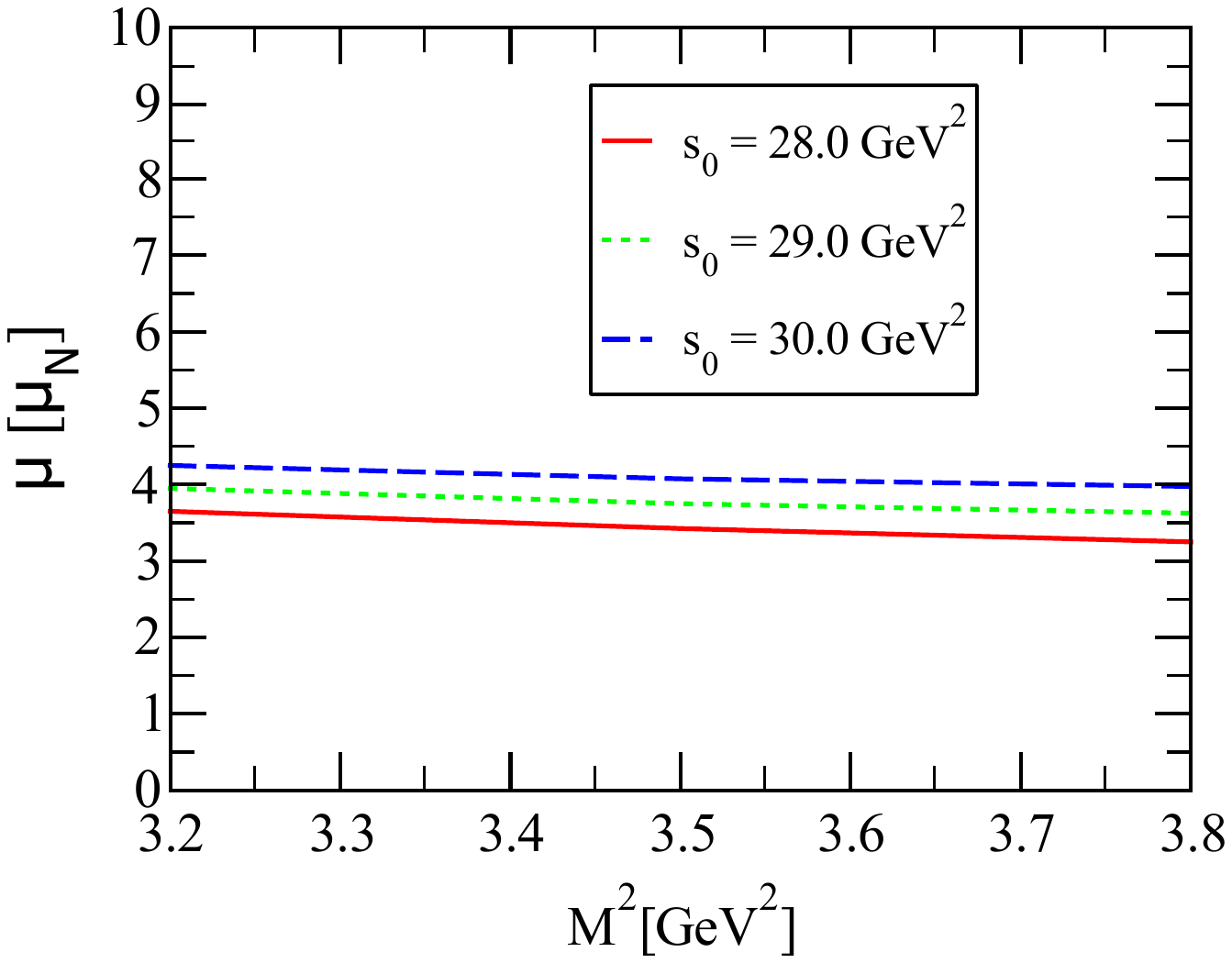}} ~~~
  \subfloat[]{\includegraphics[width=0.47\textwidth]{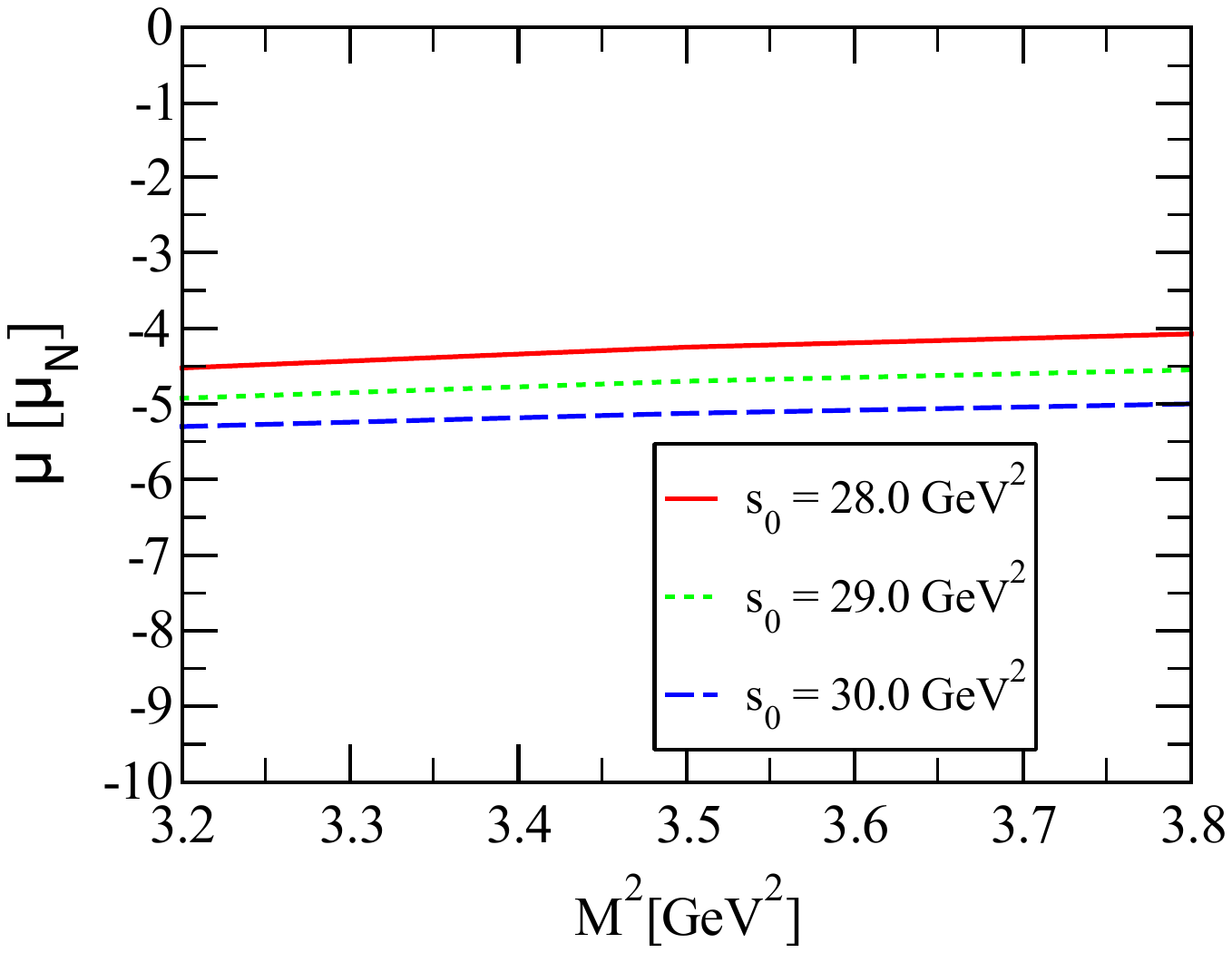}} \\
  \subfloat[]{\includegraphics[width=0.47\textwidth]{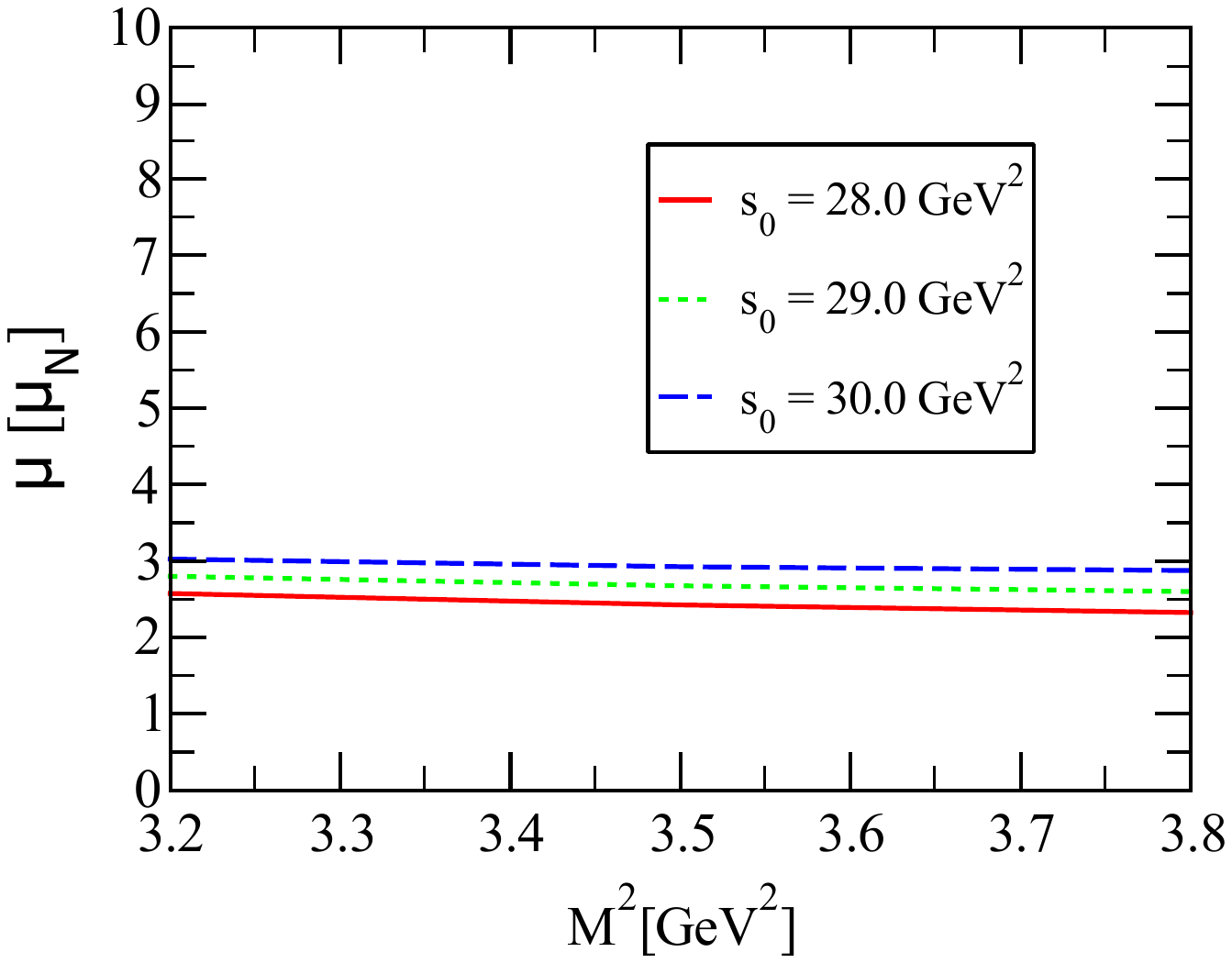}} ~~~
  \subfloat[]{\includegraphics[width=0.47\textwidth]{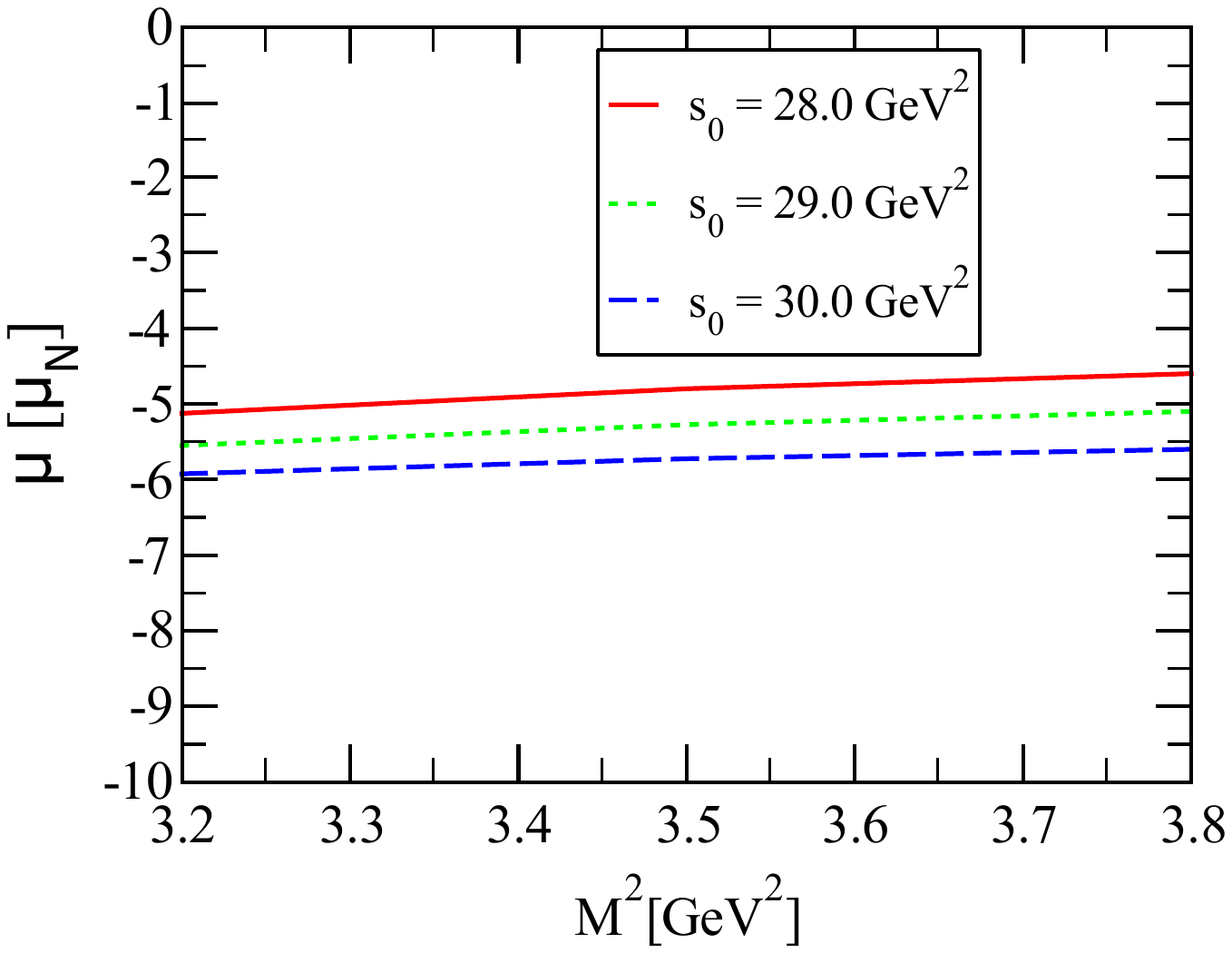}}
  \caption{ Variation of magnetic moments of the $[uc][\bar c \bar d]$ tetraquark as a function of the $\rm{M^2}$ at different values of $\rm{s_0}$: (a) for the $J_\alpha^1$ interpolating current, (b) for the $J_\alpha^2$ interpolating current, (c) for the $J_\alpha^3$ interpolating current; and (d)  for the $J_\alpha^4$ interpolating current, respectively.}
 \label{figMsq}
  \end{figure}

    \begin{figure}[htp]
\centering
  \subfloat[]{\includegraphics[width=0.47\textwidth]{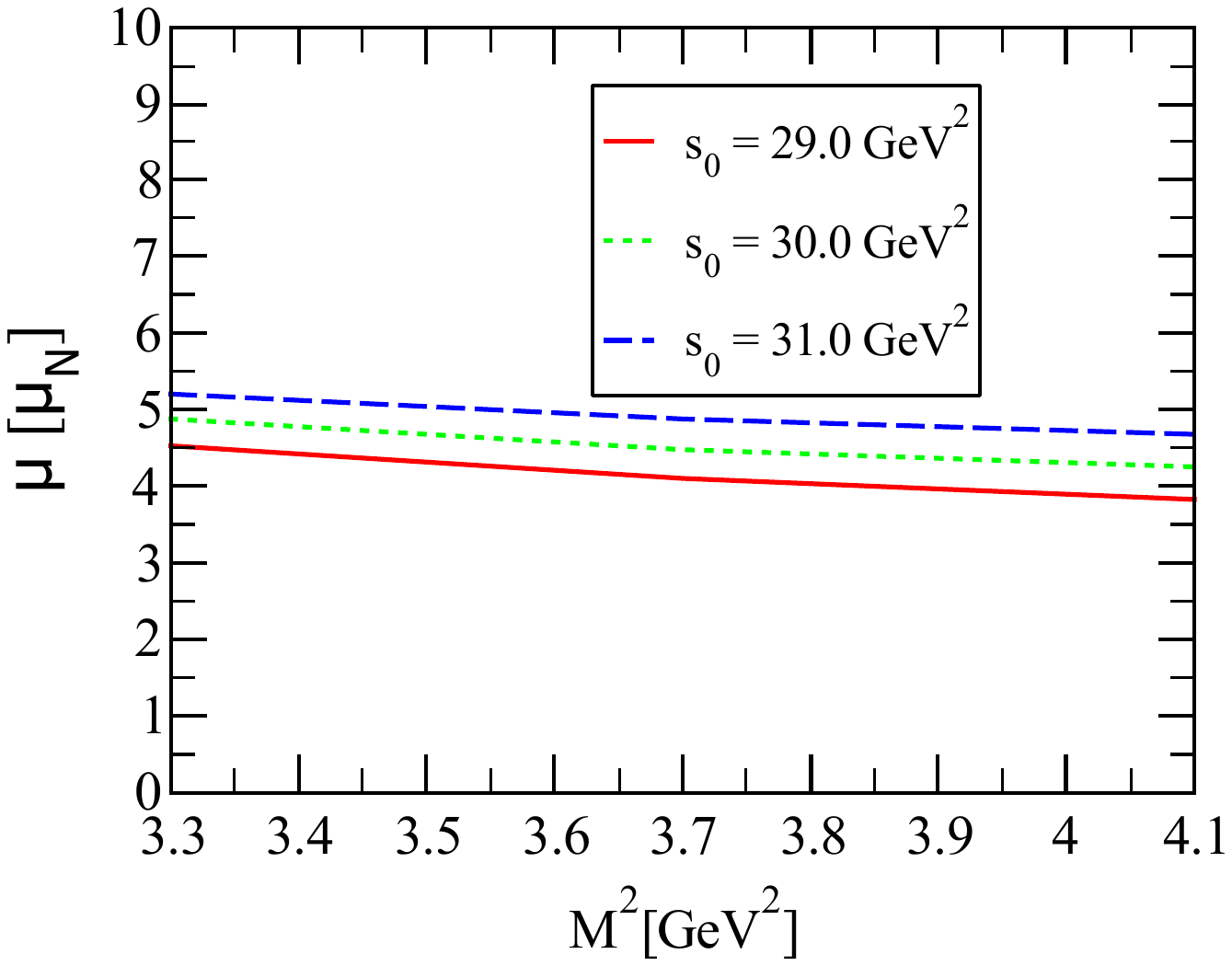}} ~~~
  \subfloat[]{\includegraphics[width=0.47\textwidth]{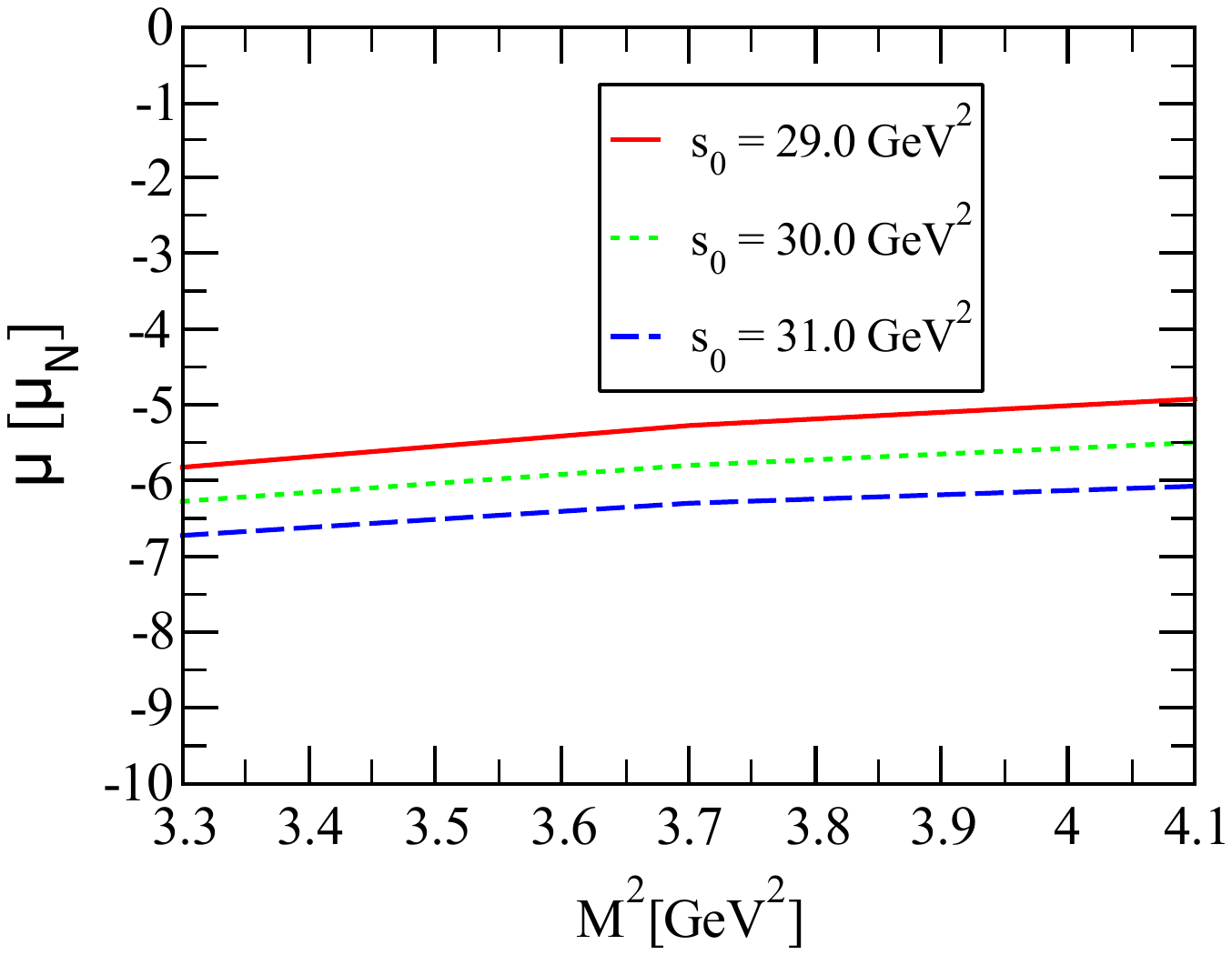}} \\
  \subfloat[]{\includegraphics[width=0.47\textwidth]{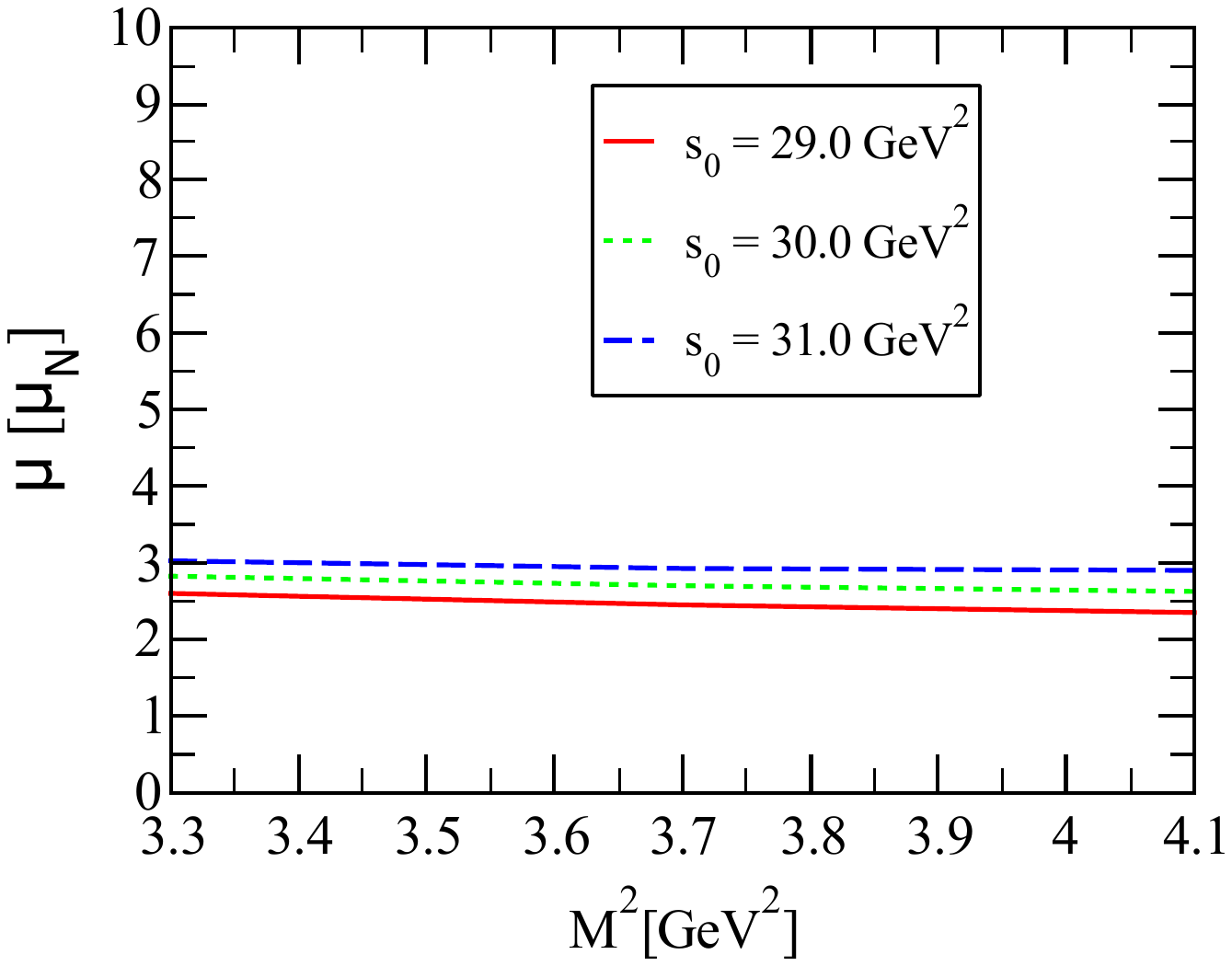}} ~~~
  \subfloat[]{\includegraphics[width=0.47\textwidth]{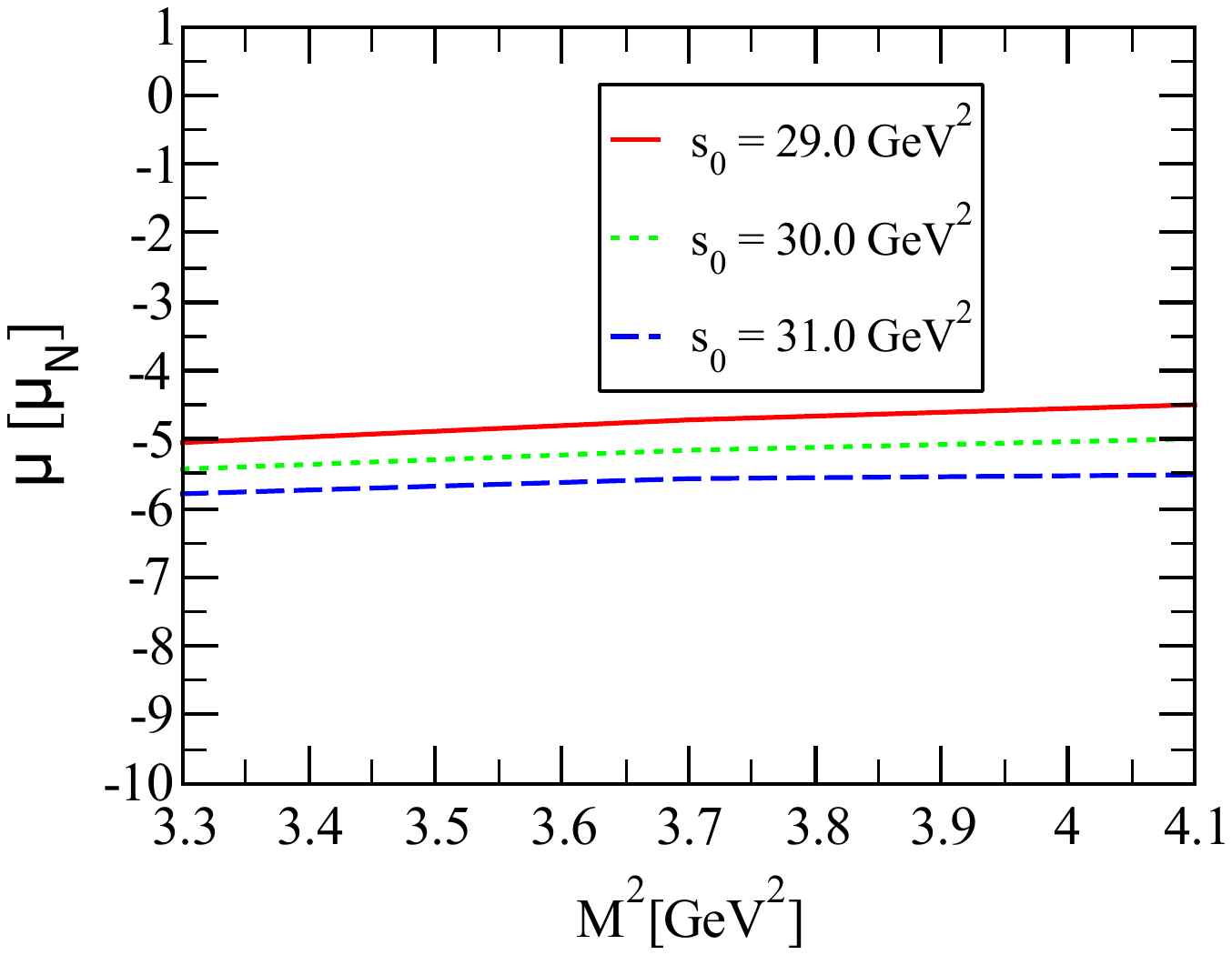}}
  \caption{ Variation of magnetic moments of the $[uc][\bar c \bar s]$ tetraquark as a function of the $\rm{M^2}$ at different values of $\rm{s_0}$: (a) for the $J_\alpha^1$ interpolating current, (b) for the $J_\alpha^2$ interpolating current, (c) for the $J_\alpha^3$ interpolating current; and (d)  for the $J_\alpha^4$ interpolating current, respectively.}
 \label{figMsq11}
  \end{figure}

     \begin{figure}[htp]
\centering
  \subfloat[]{\includegraphics[width=0.47\textwidth]{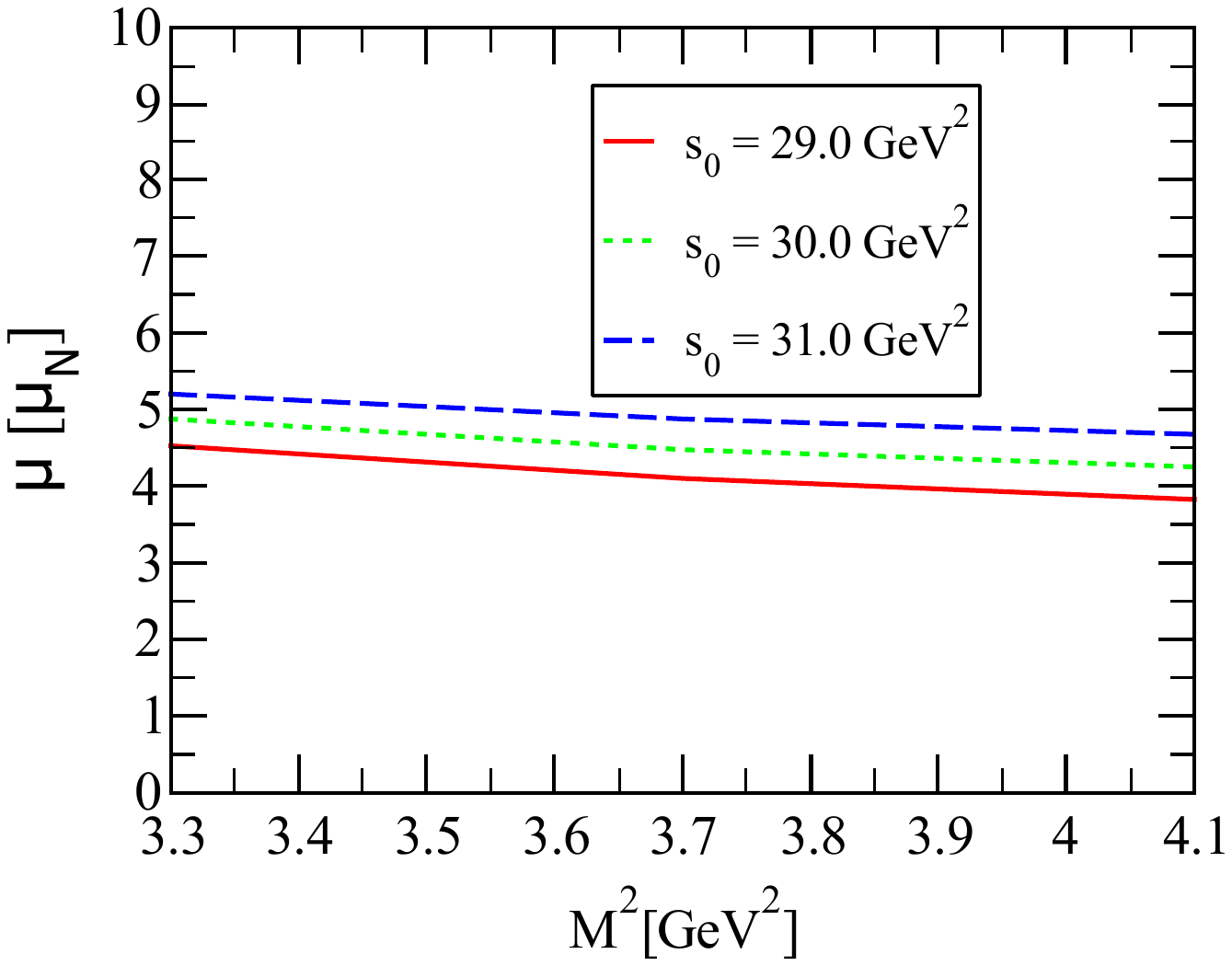}} ~~~
  \subfloat[]{\includegraphics[width=0.47\textwidth]{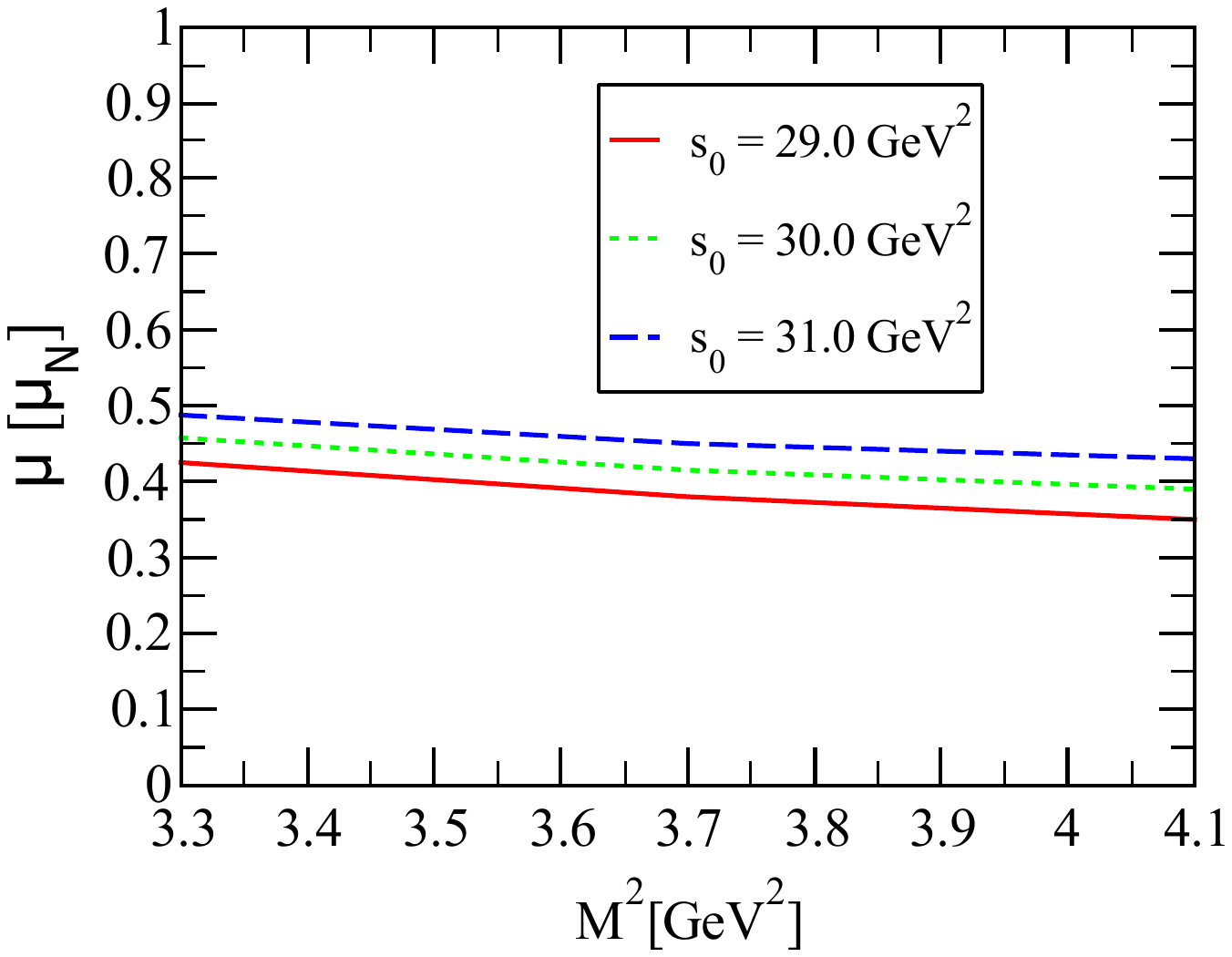}} \\
  \subfloat[]{\includegraphics[width=0.47\textwidth]{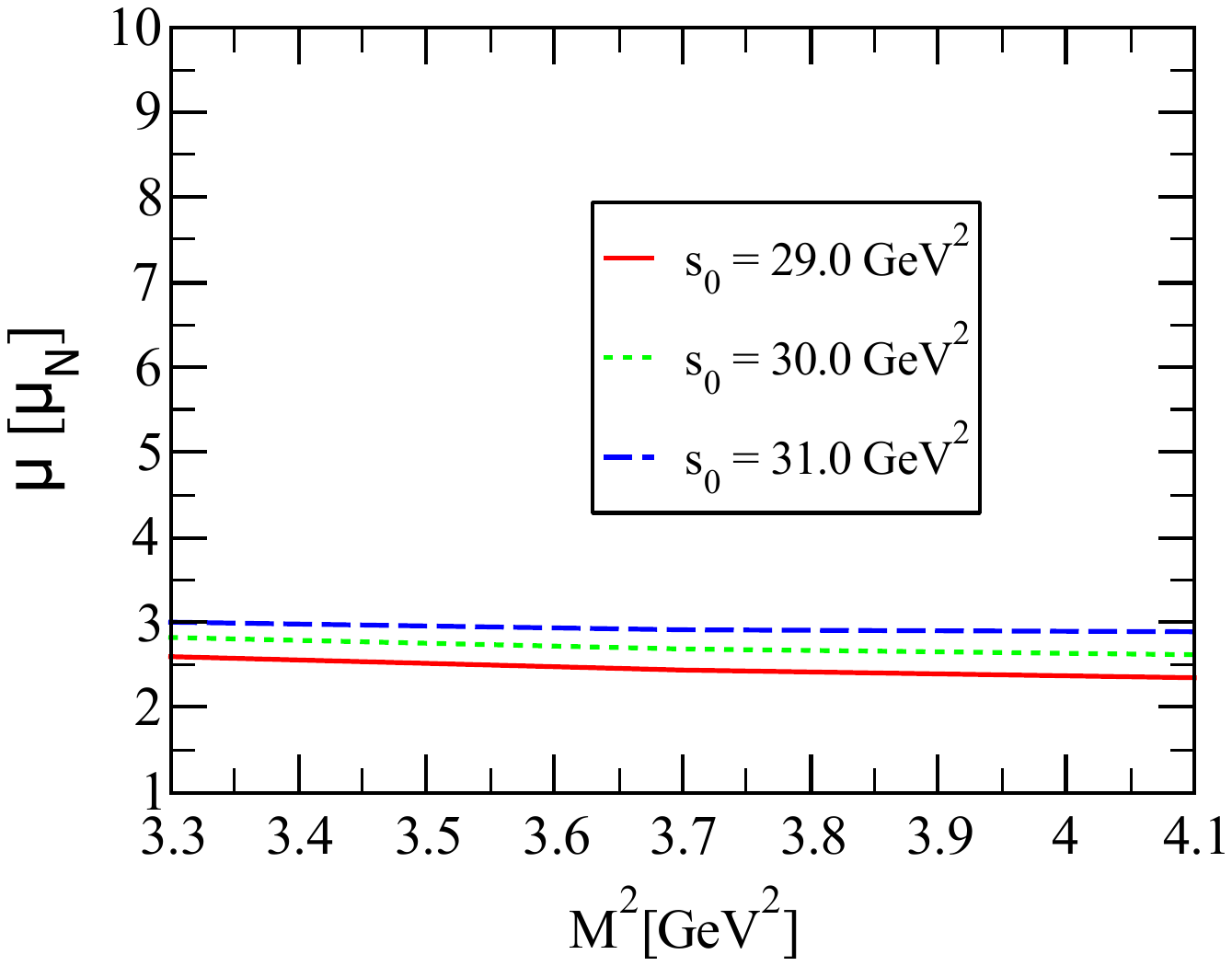}} ~~~
  \subfloat[]{\includegraphics[width=0.47\textwidth]{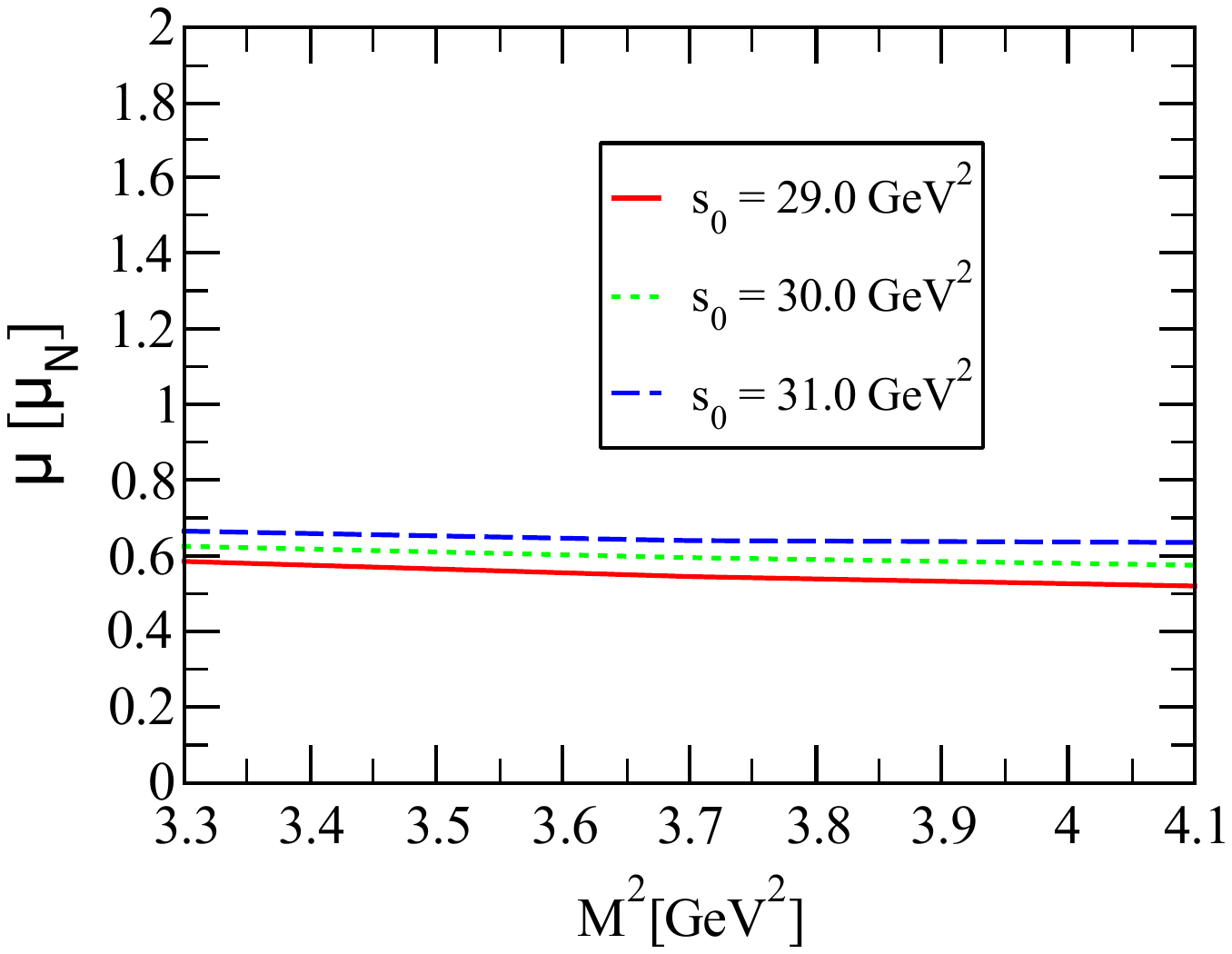}}
  \caption{ Variation of magnetic moments of the $[dc][\bar c \bar s]$ tetraquark as a function of the $\rm{M^2}$ at different values of $\rm{s_0}$: (a) for the $J_\alpha^1$ interpolating current, (b) for the $J_\alpha^2$ interpolating current, (c) for the $J_\alpha^3$ interpolating current; and (d)  for the $J_\alpha^4$ interpolating current, respectively.}
 \label{figMsq13}
  \end{figure}
  
       \begin{figure}[htp]
\centering
  \subfloat[]{\includegraphics[width=0.47\textwidth]{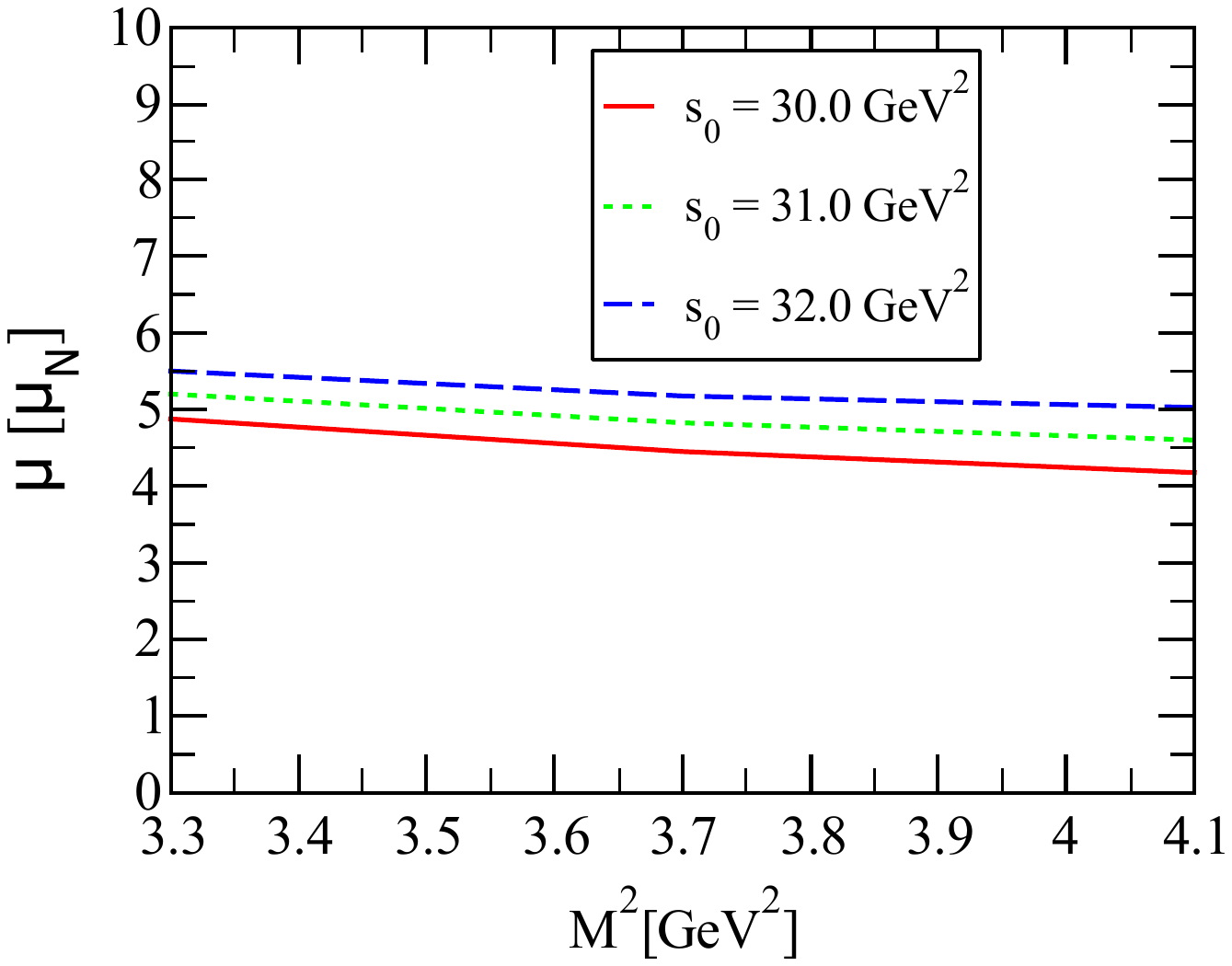}} ~~~
  \subfloat[]{\includegraphics[width=0.47\textwidth]{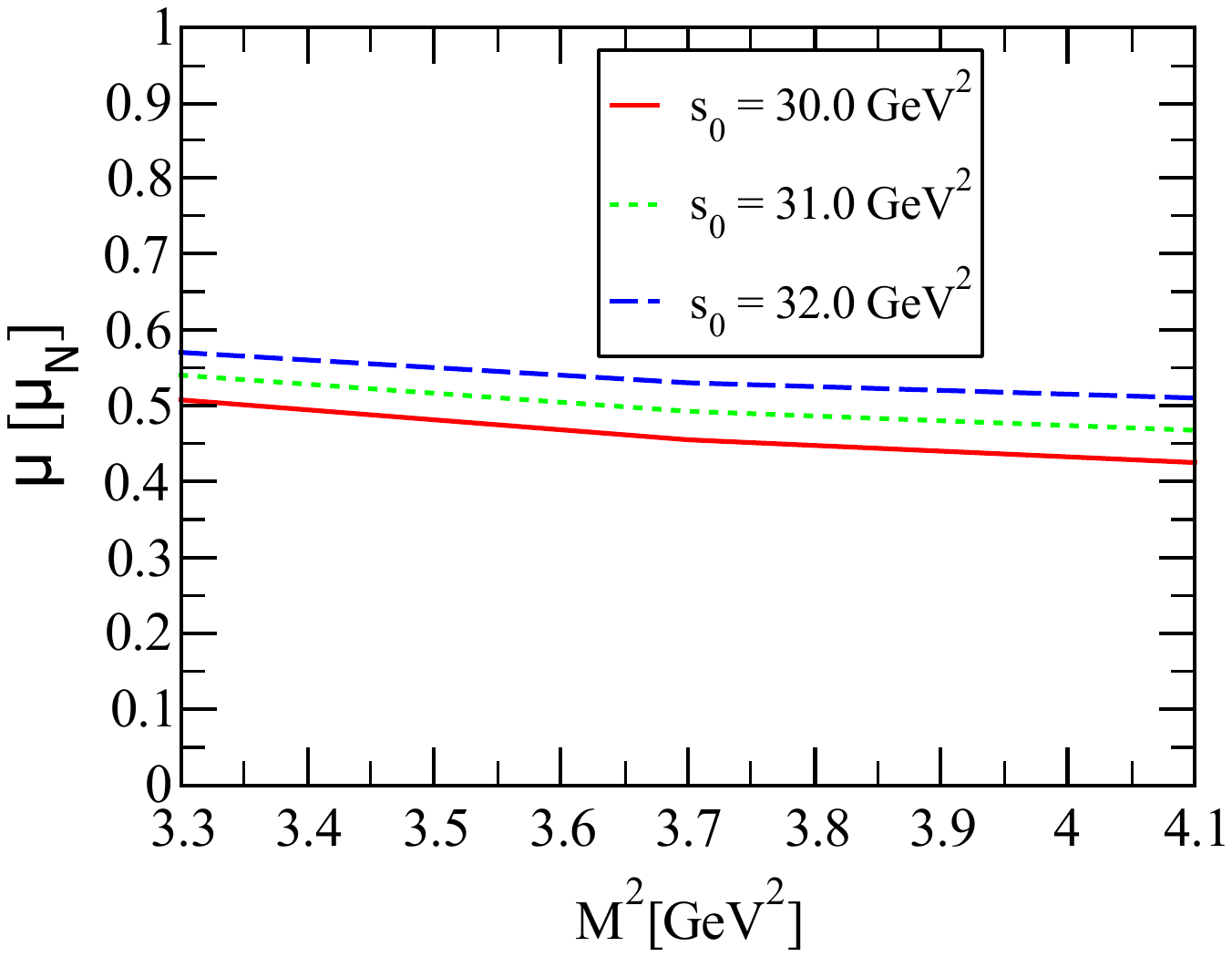}} \\
  \subfloat[]{\includegraphics[width=0.47\textwidth]{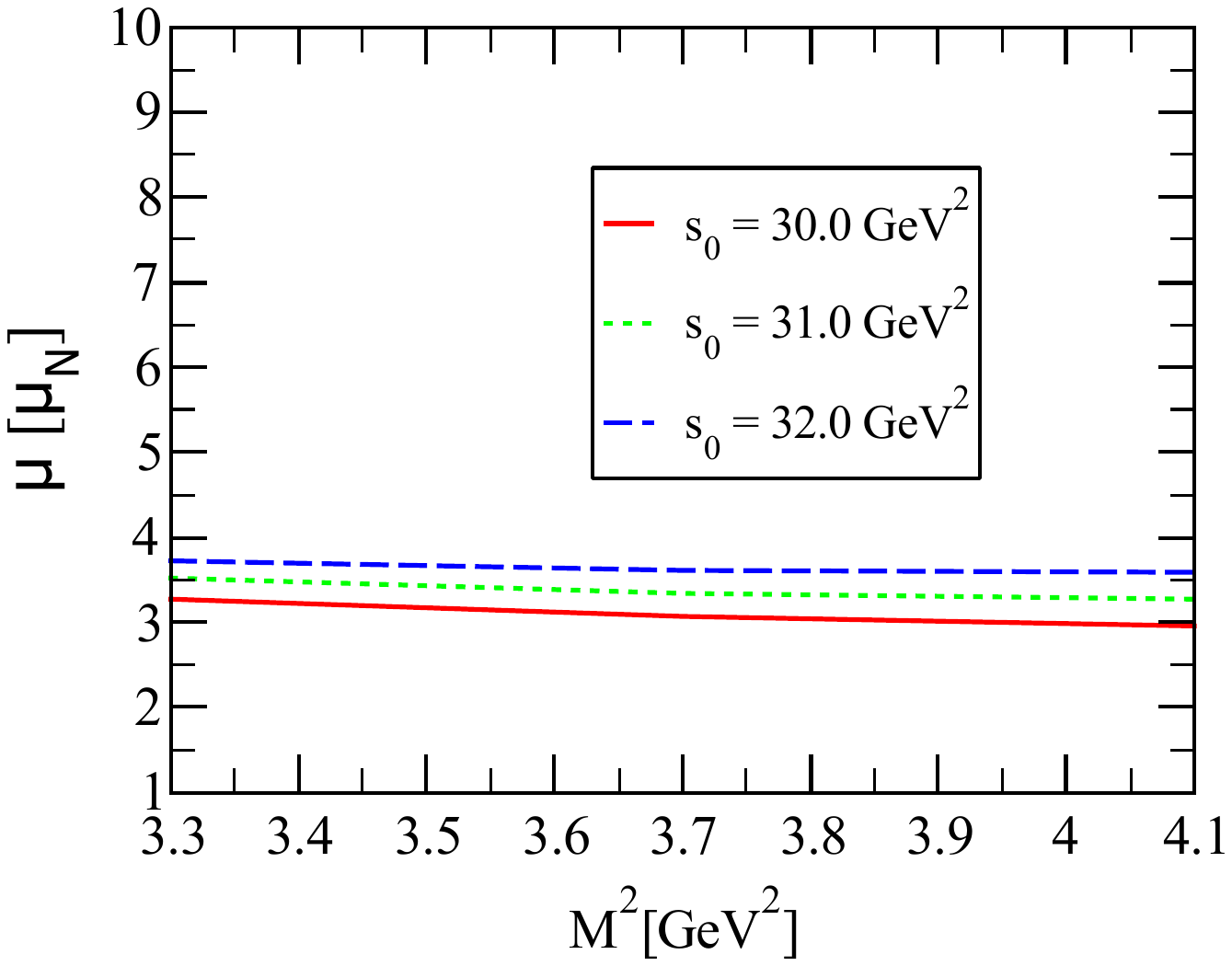}} ~~~
  \subfloat[]{\includegraphics[width=0.47\textwidth]{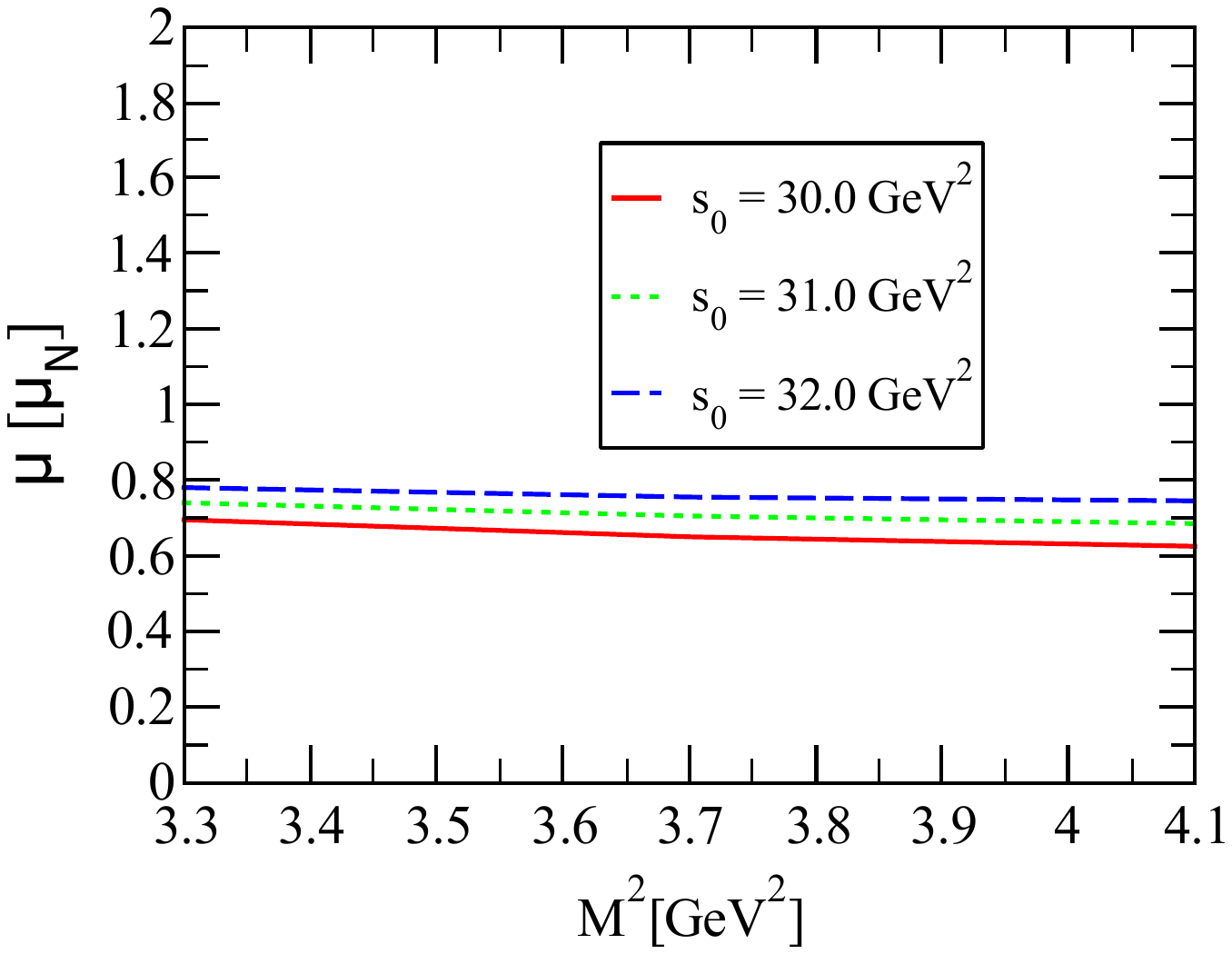}}
  \caption{ Variation of magnetic moments of the $[sc][\bar c \bar s]$ tetraquark as a function of the $\rm{M^2}$ at different values of $\rm{s_0}$: (a) for the $J_\alpha^1$ interpolating current, (b) for the $J_\alpha^2$ interpolating current, (c) for the $J_\alpha^3$ interpolating current; and (d)  for the $J_\alpha^4$ interpolating current, respectively.}
 \label{figMsq14}
  \end{figure}

 \begin{figure}[htp]
\centering
   \subfloat[]{\includegraphics[width=0.47\textwidth]{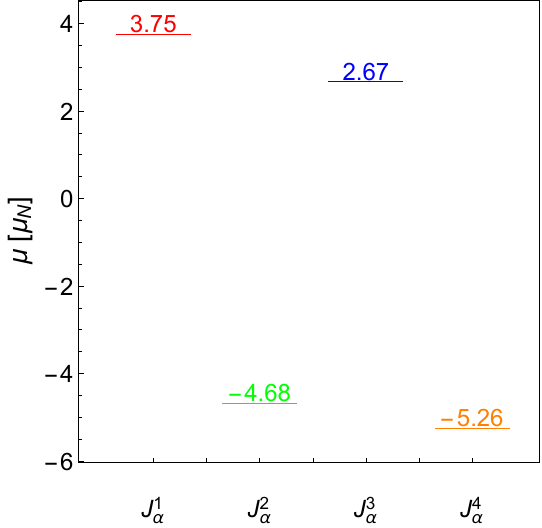}}~~~~~
  \subfloat[]{\includegraphics[width=0.47\textwidth]{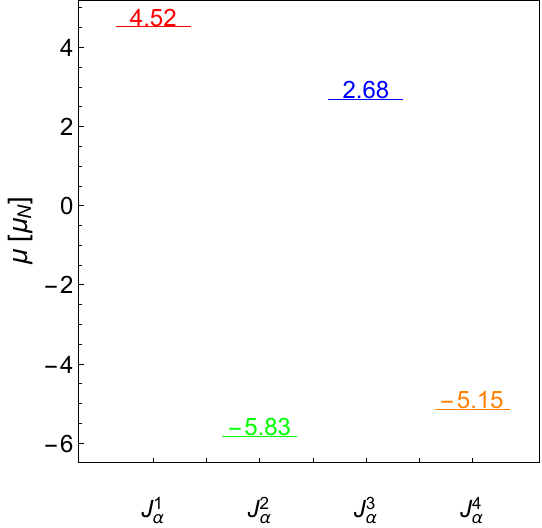}} \\
  \vspace{1 cm}
   \subfloat[]{\includegraphics[width=0.47\textwidth]{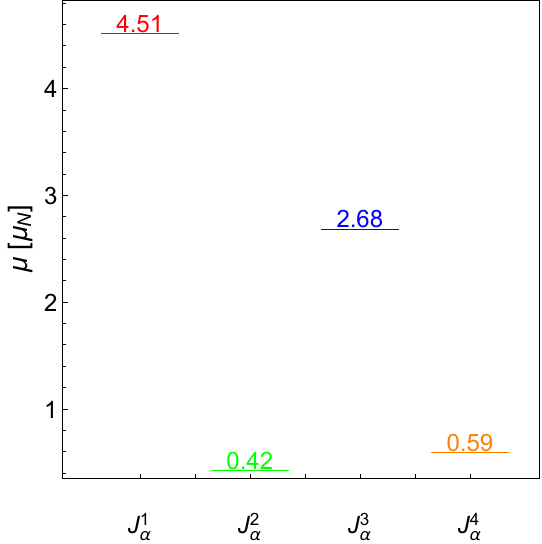}} ~~~~~
  \subfloat[]{\includegraphics[width=0.47\textwidth]{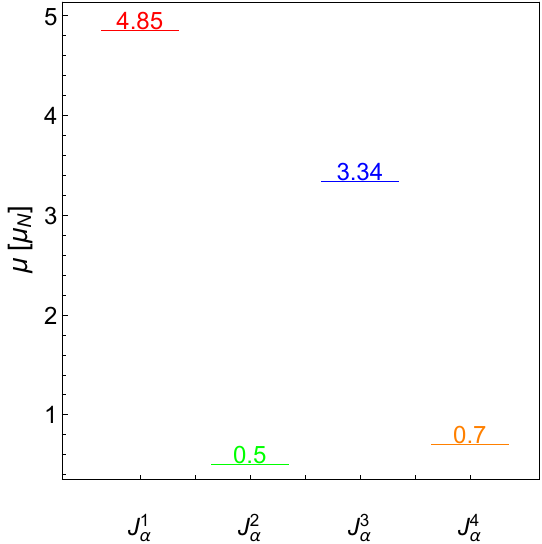}} 
  \caption{ The magnetic moments of vector hidden-charm tetraquark states for central values: (a) for $[uc][\bar c \bar d]$ states, (b)  for $[uc][\bar c \bar s]$ states, (c) for $[dc][\bar c \bar s]$ states; and (d)  for $[sc][\bar c \bar s]$ states, respectively.}
 \label{figMsq1}
  \end{figure}


  \begin{figure}[htp]
\centering
   \subfloat[]{\includegraphics[width=0.47\textwidth]{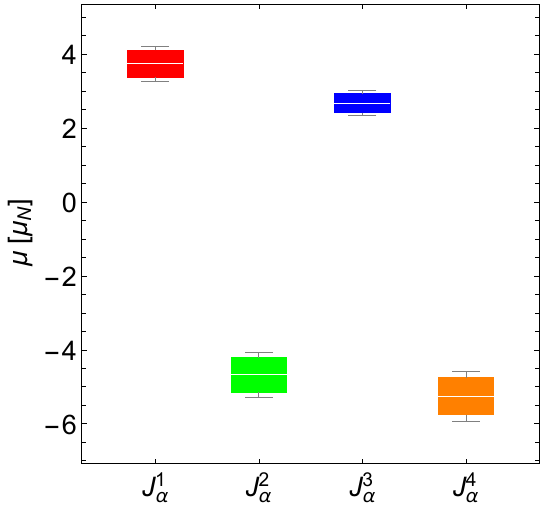}}~~~~~
  \subfloat[]{\includegraphics[width=0.47\textwidth]{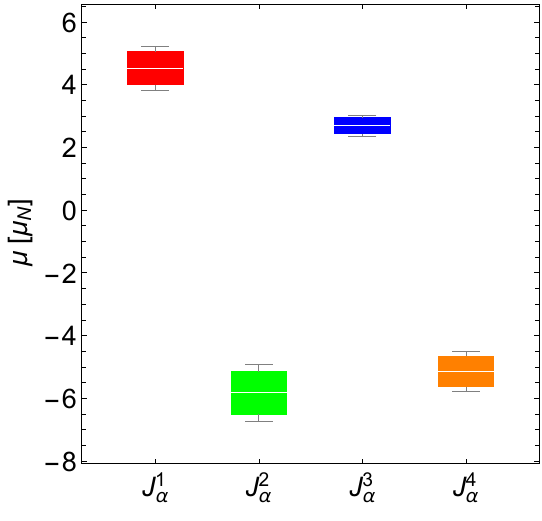}} \\
   \subfloat[]{\includegraphics[width=0.47\textwidth]{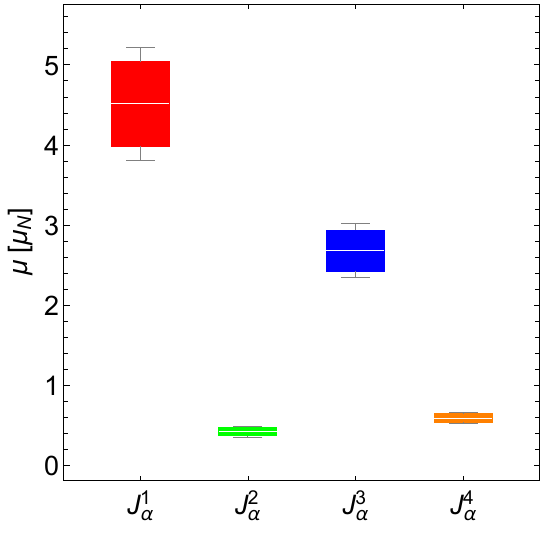}} ~~~~~
  \subfloat[]{\includegraphics[width=0.47\textwidth]{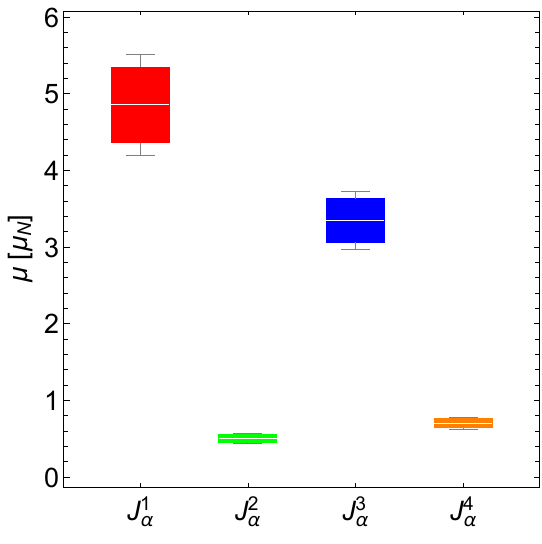}} 
  \caption{ The magnetic moments of vector hidden-charm tetraquark states for combined with errors: (a) for $[uc][\bar c \bar d]$ states, (b)  for $[uc][\bar c \bar s]$ states, (c) for $[dc][\bar c \bar s]$ states; and (d)  for $[sc][\bar c \bar s]$ states, respectively.}
 \label{figMsq2}
  \end{figure}
 
 \end{widetext}


\bibliography{Vectorhiddencharm.bib}
\bibliographystyle{elsarticle-num}

\end{document}